\documentclass[11pt,a4paper]{article}

\usepackage[margin=25mm]{geometry}
\usepackage{amsmath,amssymb}
\usepackage{graphicx}
\usepackage{booktabs}
\usepackage{array}
\usepackage{xcolor}
\usepackage{microtype}
\usepackage[font=small,labelfont=bf]{caption}
\usepackage{tikz}
\usetikzlibrary{arrows.meta,positioning,fit,calc}
\usepackage[numbers,sort&compress]{natbib}
\usepackage[colorlinks=true,linkcolor=blue!60!black,citecolor=blue!60!black,urlcolor=blue!60!black]{hyperref}
\usepackage{url}
\usepackage{placeins}          

\usepackage{longtable}         

\graphicspath{{figures/}}
\newcommand{\emgforge}{\texttt{emgforge}}
\newcommand{\dd}{\mathrm{d}}
\newcommand{\Vm}{V_{\mathrm{m}}}
\newcommand{\repo}{\url{https://github.com/dhalatsis/emgforge}}
\newcommand{\figorbox}[2][]{\IfFileExists{figures/#2.pdf}{\includegraphics[#1]{#2}}{\fbox{\parbox[c][55mm][c]{0.92\textwidth}{\centering\ttfamily missing figure: \detokenize{#2}}}}}

\title{\textbf{emgforge}: an automated end-to-end pipeline for simulating surface EMG on MRI-based volume conductors}

\author{Dimitrios Halatsis$^{1,*}$ \quad Noura Ezaz-Nikpay$^{2}$ \quad Pranav Mamidanna$^{1,3}$ \quad Dario Farina$^{1}$\\[6pt]
  \small $^{1}$Department of Bioengineering, Imperial College London, London, UK\\
  \small $^{2}$Department of Computing, Imperial College London, London, UK\\
  \small $^{3}$I-X Centre for AI in Science, Imperial College London, London, UK\\[4pt]
  \small $^{*}$Correspondence: \texttt{d.chalatsis22@imperial.ac.uk}. Code: \repo}
\date{Preprint --- September 2026}

\begin{document}
\maketitle

\begin{abstract}
Simulated electromyograms are used to understand what an electrode records, to test
decomposition and estimation algorithms on signals with known ground truth, and to train
learning-based decoders. Most simulators fix the geometry to a cylinder or a slab, or stop
at the lead field and leave the rest to the user. We present \emgforge{}, an open pipeline
that takes a labelled MRI segmentation of a limb to surface EMG in one command: a
tetrahedral mesh with conductivity tensors aligned with each muscle's fibres; one
reciprocal finite-element solve per electrode, valid for every fibre of every muscle;
fibre beds that are straight or follow the muscle's shape; a motor-unit pool obeying the
size principle; motor-unit action potentials on any electrode layout; and a
motoneuron-pool and twitch layer that turns a drive or a movement into interference EMG
and force. We describe the pipeline stage by stage, and at each stage we show, with an
example, why the choice was made. The step that turns a lead field into a single-fibre
action potential --- direct line-source synthesis --- is checked against a closed-form
solution ($r = 1.0000$, zero lag), and the whole chain against fifty checks with numeric
criteria from the physiological literature. We then use the pipeline for four studies: what
an electrode sees as a function of depth, fat, spacing and montage; whether fibre
geometry changes the signal; how much crosstalk a grid over one muscle receives from its
neighbours; and how interference EMG scales with drive. The code, the validation suite and
three datasets with ground truth are released.
\end{abstract}

\section{Introduction}
\label{sec:intro}

A forward model of the electromyogram (EMG) simulates the signal an electrode records
when muscle fibres fire. Such models are used to understand what shapes a motor-unit
action potential (MUAP), to test decomposition and estimation algorithms on signals whose
ground truth is known, and, increasingly, to generate labelled training data for
learning-based EMG decoding \citep{maksymenko2023myoelectric, ma2025conditional,
ma2024neuromotion}. The classical models are analytical: a fibre in an infinite
anisotropic medium \citep{rosenfalck1969intra, andreassen1981relationship,
nandedkar1983simulation}, a planar layered conductor \citep{farina1999compensation,
farina2001novel}, or a multilayer cylinder \citep{gootzen1991finite, blok2002three,
farina2004multilayer}. Finite-element models lift the geometric restrictions
\citep{lowery2002multiple, lowery2004volume, botelho2019anatomically}, and with
reciprocity they run on MRI-derived anatomy at interactive speed: the myoelectric digital
twin of \citet{maksymenko2023myoelectric} is the reference example, and generative
surrogates trained on such twins \citep{ma2025conditional, ma2024neuromotion} make the
output cheap. Open code exists at the planar level \citep{petersen2019comprehensive} and
as multi-domain finite-element frameworks \citep{klotz2020modelling}. What has been
missing is an open pipeline that goes from a segmentation to EMG without hand steps, on
any geometry a segmentation describes, with every stage exposed so it can be inspected,
replaced and tested.

\emgforge{} is that pipeline. Its input is a labelled segmentation (muscles, bones, fat
and skin) and an electrode layout; its output is EMG on those electrodes, with the
spike trains, motor-unit properties, lead fields and MUAPs that produced it. The stages
are the ones every volume-conductor model has --- mesh, conductivities, lead fields,
fibres, motor units, action potentials, activation --- but each is a checkable object with
a small interface, so the same activation layer runs on an analytical cylinder, a
parametric limb or an MRI forearm, and the same validation checks run on all three. We
built it because three of our own efforts needed it: a decomposition benchmark whose
synthetic ground truth must be inspectable \citep{mamidanna2025muniverse}, a
model-informed decomposition that inverts a forward model and is only as good as that
model \citep{halatsis2025bmiss}, and neural-operator surrogates of the lead field that
inherit whatever the finite-element stage gets wrong \citep{halatsis2024modelling,
halatsis2026neural}.

The paper is written to be read in order. Section~\ref{sec:overview} gives the whole
chain in one figure and one table of runtimes. Section~\ref{sec:steps} walks through it
stage by stage; at each stage we state the decision we took and show, with an example, why.
Section~\ref{sec:validation} is the validation: the synthesis step against a closed-form
solution, the finite-element conductor against the analytical cylinder, and the MUAPs and
interference signals against the quantitative literature. Section~\ref{sec:studies} uses
the pipeline for four studies and states what each one teaches. Section~\ref{sec:discussion}
discusses what the pipeline is for, what it does not yet do, and what comes next.

The contributions are: (i) an automated, anatomy-general pipeline from segmentation to
EMG with one entry point and a per-stage cost that scales with electrodes, not with
fibres or contractions; (ii) a synthesis method, \emph{direct line-source synthesis},
that reproduces a closed-form line-source solution exactly and stays usable on
finite-element lead fields; (iii) a reusable validation suite of fifty literature-based
checks; (iv) four studies with conclusions for people who design electrodes, choose fibre
models or interpret interference EMG; (v) three released datasets with ground truth.

\section{The pipeline at a glance}
\label{sec:overview}

\begin{figure}[tbp]
\centering
\resizebox{\textwidth}{!}{%
\begin{tikzpicture}[
  node distance=4mm and 6mm,
  box/.style={draw, rounded corners=2pt, minimum height=9mm, text width=31mm, align=center, font=\small, fill=gray!5},
  key/.style={box, fill=green!8, draw=green!50!black},
  arr/.style={-{Latex[length=2mm]}, thick}]
\node[box] (seg) {segmentation\\ \scriptsize labels: muscles, bone, fat, skin};
\node[box, right=of seg] (mesh) {mesh + tensors\\ \scriptsize Gmsh; fibre-aligned $\boldsymbol\sigma$};
\node[box, right=of mesh] (lf) {lead fields\\ \scriptsize one FEM solve per electrode};
\node[key, right=of lf] (sfap) {SFAP synthesis\\ \scriptsize direct line-source};
\node[box, below=of seg] (fib) {fibre beds\\ \scriptsize straight / streamline};
\node[box, right=of fib] (mu) {motor-unit pool\\ \scriptsize size principle};
\node[box, right=of mu] (muap) {MUAPs\\ \scriptsize any electrode layout};
\node[box, right=of muap] (act) {activation\\ \scriptsize pool, twitch, drive};
\node[box, right=of act] (emg) {EMG \& force\\ \scriptsize static / movement};
\draw[arr] (seg) -- (mesh); \draw[arr] (mesh) -- (lf); \draw[arr] (lf) -- (sfap);
\draw[arr] (seg) -- (fib); \draw[arr] (fib) -- (mu); \draw[arr] (mu) -- (muap);
\draw[arr] (sfap) -- (muap); \draw[arr] (muap) -- (act); \draw[arr] (act) -- (emg);
\draw[arr] (fib) -- (lf);
\end{tikzpicture}}
\caption{The \emgforge{} chain. The lead field is the hinge: computed once per electrode
and sampled along every fibre of every muscle; direct line-source synthesis turns each
sampled lead field into an SFAP; fibres are summed per motor unit, and the activation
layer turns spike trains into interference EMG and force.}
\label{fig:pipeline}
\end{figure}
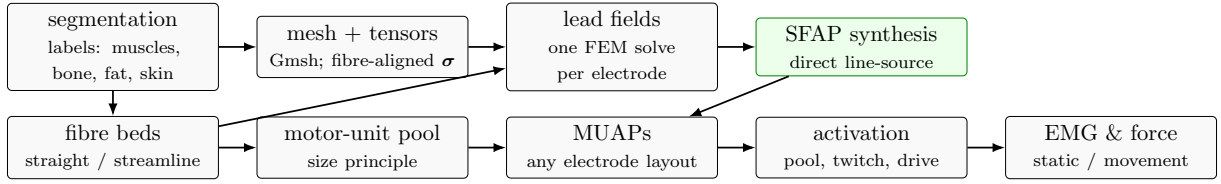

Figure~\ref{fig:pipeline} shows the chain. A segmentation is meshed and given
conductivities (Sec.~\ref{sec:anatomy}); one finite-element solve per electrode gives that
electrode's lead field over the whole limb (Sec.~\ref{sec:leadfield}); fibres are laid
through each muscle (Sec.~\ref{sec:fibres}) and grouped into motor units
(Sec.~\ref{sec:pool}); the lead field sampled along each fibre is turned into a
single-fibre action potential (SFAP) and the fibres of a unit are summed into its MUAP on
every electrode (Sec.~\ref{sec:direct}); and a motoneuron pool driven by an excitation or
a joint angle turns the MUAPs into interference EMG and force
(Sec.~\ref{sec:activation}). Everything downstream of the solve is NumPy; only the solve
needs FEniCSx \citep{baratta2023dolfinx} and Gmsh \citep{geuzaine2009gmsh}.

\paragraph{One command.} The whole chain runs cold from the committed segmentation with
\texttt{python scripts/run\_pipeline.py --n-mu 100} (defaults: the FCU, a $5\times5$ grid
at $10$~mm; any muscle label, grid and spacing can be given), which writes the mesh, the
lead fields, the pool, the MUAP tensor and a contraction with its spike trains, force
and EMG, together with the wall time of each stage. Table~\ref{tab:pipeline} gives those
times for the forearm used throughout this paper (subject WR, flexor carpi ulnaris, a
$5\times5$ grid). The point of the table is the shape of the cost: the finite-element work
is per electrode and is done once; fibres, units and contractions reuse it.

\begin{table}[tbp]
\centering\footnotesize
\caption{One cold run of the pipeline from the committed forearm segmentation to
interference EMG on the grid (subject WR, FCU, $5\times5$ grid, $100$ units, two worker
processes on a desktop CPU shared with other jobs): what each stage does, its input and
output sizes and its wall time. The finite-element work (mesh, tensors, $25$ lead
fields) is done in under a minute and once; the recipe's monopole fits and the MUAP
synthesis are per electrode--fibre pair and per unit and dominate; a contraction costs
seconds.}
\label{tab:pipeline}
\IfFileExists{pipeline_table.tex}{
\begin{tabular}{@{}l >{\raggedright\arraybackslash}p{5.6cm} >{\raggedright\arraybackslash}p{3.7cm} r@{}}
\toprule
Stage & What it does & Input $\rightarrow$ output & Wall (s) \\
\midrule
Mesh & resample the labels, marching cubes on the tissue mask, smooth and decimate the surface, fTetWild tetrahedra, tissue tag per cell from the label under its centroid & 320$\times$320$\times$60 voxels $\rightarrow$ 46\,749 tetrahedra, 11\,022 nodes & 42.4 \\
Fibre directions & PCA fibre direction and smoothed per-slice centreline of every muscle label $\rightarrow$ the fibre-aligned (5:1) conductivity tensor of each muscle & 21 labels $\rightarrow$ 17 muscles, 17 centrelines & 2.2 \\
Muscle geometry & ray-cast boundary radius of each muscle cross-section per slice $\rightarrow$ the morphing-disk frame that carries a fibre through the muscle & 17 centrelines $\rightarrow$ 17 cross-section models & 6.0 \\
Fibre bed & Poisson-disk sampling of the muscle cross-section at the set density; one curved morphing-disk path per fibre from tendon to tendon & FlexorCarpiUlnaris, 206 mm$^2$ $\rightarrow$ 637 fibres $\times$ 200 points (204 mm long) & 1.3 \\
Motor-unit pool & exponential Henneman size distribution; each unit's territory grown around a random anchor fibre until it holds its fibres; index = size rank = recruitment order & 637 fibres $\rightarrow$ 100 units of 5--395 fibres & 0.0 \\
Volume conductor & conductivity tensor per cell from its tissue tag (muscle tensors rotated onto the centreline tangent), a skin shell inside the fat, FEniCSx spaces and search trees & 46\,749 cells $\rightarrow$ 46\,749 $\sigma$ tensors, 11\,022 dofs & 5.3 \\
Electrode grid & M$\times$N grid ray-cast onto the skin surface over the muscle: rows IED apart along the arm, columns IED apart along the skin arc, centred on the limb-axis $\rightarrow$ muscle ray & 11\,188 skin triangles $\rightarrow$ 25 electrodes (5$\times$5, 10.1$\times$10.0 mm) & 3.4 \\
Lead fields & one reciprocity solve per electrode (zero-mean Gaussian source at the skin point, GMRES/ILU); $\varphi$ sampled along every fibre of the bed & 25 solves $\rightarrow$ $\varphi$ bank 25$\times$637$\times$200 & 17.8 \\
Lead-field conditioning & the recipe's denoising: a free-position 3-monopole fit of each $\varphi$(z) that removes mesh ripple before the second derivative --- once per electrode--fibre pair & 25$\times$637 $\varphi$(z) $\rightarrow$ 15\,925 monopole fits & 253.5 \\
MUAPs & per unit and electrode: line-source integral of the current-source density of a Rosenfalck action potential (both directions from the NMJ, one-sided tendon window) against $\varphi$ over the unit's fibres; physical time, t = 0 at the NMJ & $\varphi$ + 100 units $\rightarrow$ MUAP tensor 100$\times$25$\times$256 (280\,175 SFAPs) & 658.6 \\
Activation & motoneuron pool (recruitment thresholds, onion-skin rate coding, renewal ISIs) $\rightarrow$ spikes; twitch model $\rightarrow$ force; spikes convolved with the MUAPs $\rightarrow$ EMG on the grid, one trapezoid per drive level & 6 trapezoids $\times$ 3.2 s $\rightarrow$ EMG 25$\times$6554 per level, force, 32\,858 spikes & 1.8 \\
\midrule
Total & sum of the stages above & & 992 \\
\bottomrule
\end{tabular}
}{\emph{(pipeline\_table.tex not generated yet)}}
\end{table}

\paragraph{Why reciprocity.} By the reciprocity theorem for a linear volume conductor
\citep{malmivuo1995bioelectromagnetism}, the potential at electrode $e$ due to a unit
current source at a point $\mathbf{r}$ inside a muscle equals the potential at
$\mathbf{r}$ due to a unit current injected at $e$. So one solve with the source at the
electrode gives the \emph{lead field} $\varphi_e(\mathbf{r})$ everywhere, and sampling it
along a fibre path $\mathbf{r}(z)$ gives $\varphi_e(z)$, the weight with which membrane
current at $z$ reaches the electrode. Every fibre, every motor unit, every muscle and
every contraction reuse the same solve; only the number of electrodes costs finite-element
work. This is the design decision that makes the rest cheap, and the one that lets the
crosstalk study of Sec.~\ref{sec:study-crosstalk} use the same $25$ solves for five
muscles. The interior-point reciprocity of our solver is checked directly
(Sec.~\ref{sec:val-cylinder}).

\section{The pipeline, stage by stage}
\label{sec:steps}

\subsection{Anatomy from a segmentation}
\label{sec:anatomy}
The input is a label map: each muscle, the bones, and fat/skin. Each label is meshed
into a tetrahedral domain with a thin skin shell (Gmsh), and each domain receives a
conductivity: isotropic for fat, skin and bone, transversely isotropic for muscle with a
ratio of $5$ along the fibre \citep{gielen1984electrical, rush1963resistivity}. The muscle
tensor is rotated onto that muscle's own centreline tangent, estimated from per-slice
centroids and a smoothing spline, so every muscle is anisotropic along its own axis. The
conductivities follow \citet{gabriel1996dielectric} at $\approx100$~Hz
(Table~\ref{tab:tissue}). The forearm used here (subject WR) meshes to $\approx47{,}000$ cells in $42$~s from a
$320\times320\times60$ label volume (Table~\ref{tab:pipeline}).

\emph{Why fibre-aligned anisotropy matters.} In an infinite medium the anisotropy
elongates the lead field along the fibres by exactly $\sqrt{5}$, and in a layered
cylinder by $1.5$ (checked in Sec.~\ref{sec:validation}); a muscle whose tensor pointed
along the limb rather than along its own fibres would spread every MUAP along the wrong
direction. \emph{Why a distinct skin layer.} The literature disagrees by $2000\times$ on
skin conductivity (Table~\ref{tab:tissue}), and a thin conductive skin layer changes the
lateral decay of the potential \citep{blok2002three}; the pipeline keeps skin as its own
domain so the choice is a parameter, not a mesh rebuild.

For validation the pipeline also builds \emph{parametric limbs} --- layered discs
(cancellous and cortical bone, muscle, fat, skin) extruded or lofted along $z$, circular
or elliptical, optionally tapered --- and carries a line-by-line Python port of the
analytical four-layer cylinder of \citet{farina2004multilayer}, whose lead field is
obtained by inverting its transfer function. The validation cylinder uses the analytical
values in both models (Table~\ref{tab:tissue}).

\begin{table}[t]
\centering\small
\caption{Volume-conductor parameters (S/m). The validation cylinder uses the analytical
values in both models; the production FEM/MRI table follows \citet{gabriel1996dielectric}
at $\approx 100$~Hz.}
\label{tab:tissue}
\footnotesize\setlength{\tabcolsep}{4pt}
\begin{tabular}{lccccc>{\raggedright\arraybackslash}p{30mm}}
\toprule
 & bone & muscle $\sigma_\perp$ & muscle $\sigma_\parallel$ & fat & skin & radii (mm) \\
\midrule
analytical~\citep{farina2004multilayer} & 0.02 & 0.10 & 0.50 & 0.04 & 1.0 & 10 / 35 / 38 / 40 \\
FEM, validation cylinder & 0.02 & 0.10 & 0.50 & 0.04 & 1.0 & 9, 10 / 35 / 38 / 40; $L=240$ \\
FEM / MRI, production & 0.02$^{\dagger}$ & 0.2455 & 1.228 & 0.0379 & $4.55\times10^{-4}$ & from the segmentation \\
\bottomrule
\multicolumn{7}{p{150mm}}{$^{\dagger}$cortical; cancellous bone 0.075 S/m. The skin values differ by
$2000\times$ between the two tables: the analytical school treats skin as a thin conductive
layer ($1$~S/m), the Gabriel-based tables as dry skin ($4.6\times10^{-4}$~S/m); it is the
largest disagreement in the literature.}
\end{tabular}
\end{table}

\subsection{Lead fields by reciprocity}
\label{sec:leadfield}
For each electrode the pipeline solves the quasi-static conductivity equation
\begin{equation}
  \nabla\!\cdot\!\big(\boldsymbol{\sigma}(\mathbf{r})\nabla u\big) = -\,s(\mathbf{r}), \qquad
  (\boldsymbol{\sigma}\nabla u)\cdot\mathbf{n} = 0 \text{ on } \partial\Omega ,
\end{equation}
with the source $s$ a zero-mean Gaussian blob of width $\sigma_{\mathrm{s}}$ centred on
the electrode (its mean is subtracted so the pure-Neumann problem is solvable), by
GMRES/ILU in FEniCSx. Electrodes are placed on the skin by ray-casting a layout onto the
mesh; the layout used in this paper is a regular $5\times5$ grid with $10$~mm
inter-electrode distance over the flexor carpi ulnaris (FCU), rows along the arm and
columns around it (measured $10.2$~mm along, $10.0$~mm across). On the forearm the $25$ solves
and the sampling of every field along the $637$ fibres of the FCU bed take $18$~s in all
(Table~\ref{tab:pipeline}); on the $1.03\times10^{6}$-cell validation cylinder a solve
takes $12$~s. Figure~\ref{fig:mri}c,d shows one grid electrode's lead field on a slice and along
three fibres at increasing distance: it is a smooth peak whose width grows with distance,
the shape a sum of a few point sources has, which is what the synthesis step later
exploits.

\begin{figure}[tbp]
\centering
\figorbox[width=\textwidth]{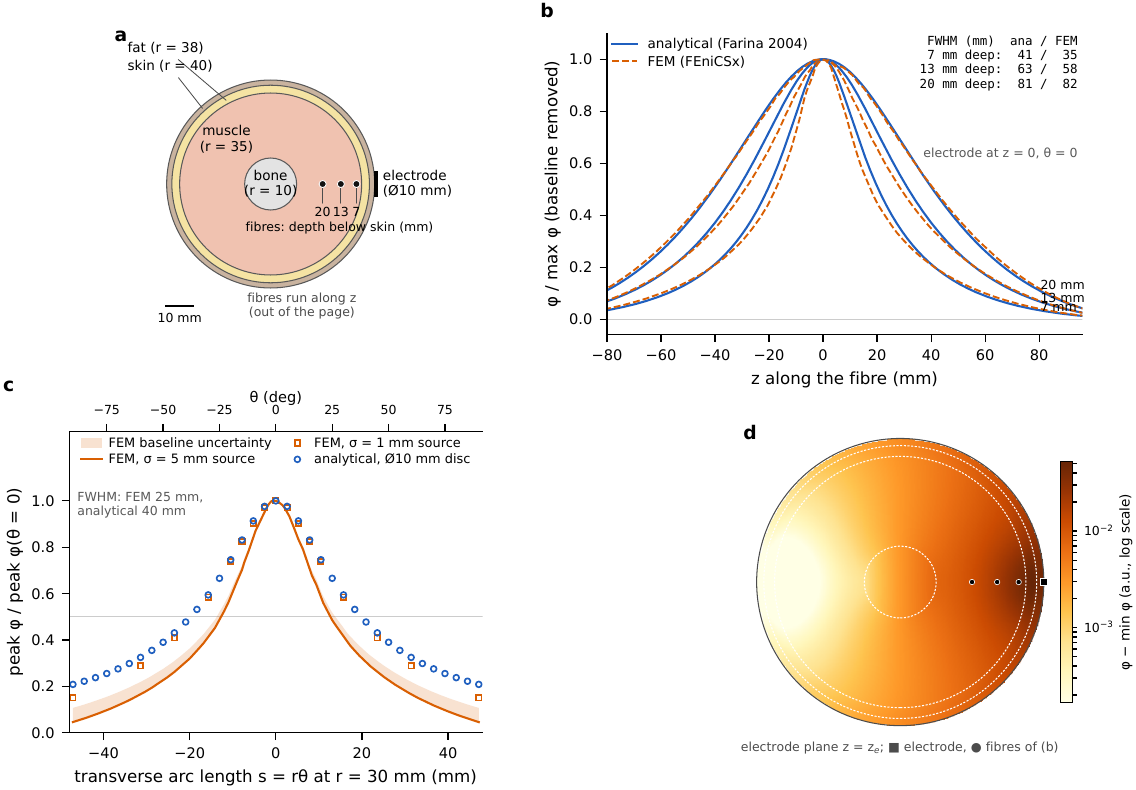}
\caption{Lead fields on the validation cylinder. (a) Cross-section with the electrode
and the fibre depths used. (b) $\varphi(z)$ along fibres 7, 13 and 20~mm below the skin:
analytical (solid) and FEM (dashed); the widths agree for deep fibres and differ for the
shallowest, where the electrode models differ (a $\varnothing10$~mm disc against a
Gaussian source that behaves as a $\varnothing5$~mm disc). (c) Transverse decay at the
fibre depth ($r=30$~mm): the FEM field around the circumference (FWHM $25$~mm; the band
is the baseline uncertainty) against the analytical peak potential for fibres displaced
around the circumference (FWHM $40$~mm); with a $\sigma_{\mathrm s}=1$~mm source the FEM
profile matches the analytical one (extra points), so the difference is the $5$~mm
electrode blob, not the conductor. (d) The FEM lead field of one electrode on the
cross-section (log scale); the dots are the fibres of (b).}
\label{fig:leadfield}
\end{figure}

\emph{Why the source width is a decision.} A Gaussian source of $\sigma_{\mathrm s} =
5$~mm extends through $2$~mm of skin and $3$~mm of fat into the muscle, and its transverse
footprint at the fibre depth is narrower than that of the analytical disc electrode
(FWHM $25$ against $40$~mm, Fig.~\ref{fig:leadfield}c); a $1$~mm source reproduces the
analytical profile. The width therefore sets the lateral footprint of every MUAP the
pipeline produces, and it should be chosen to represent the physical electrode rather
than for numerical convenience.

\subsection{Fibre geometry}
\label{sec:fibres}
Two fibre models are available for any muscle in the segmentation. The \emph{straight
bed} places fibres on the muscle's mid-length cross-section by Poisson-disk sampling at a
chosen density ($4$~fibres/mm$^2$ for the FCU) and extends them as full-length paths
parallel to the muscle centreline, with a junction at an atlas innervation-zone fraction
($0.305$ of the length for the FCU). This is the classical assumption and what cylinder
models assume. The \emph{harmonic-streamline bed} resolves the muscle's internal
architecture from its label mask alone: solve the masked Laplace problem
$\nabla^2\phi = 0$ inside the muscle with $\phi=0$ on the proximal cap, $\phi=1$ on the
distal cap and an insulated surface, and trace the streamlines of $\nabla\phi$ from seeds
on the mid-belly cross-section. Streamlines follow the muscle's shape, stay inside it and
never cross, by construction; each is one full-length fibre with a single junction at the
muscle's shared innervation zone, and a quality guard rejects tracer failures (turn per
step above $15^\circ$, or a longitudinal span below $30\%$ of the muscle). An experimental
variant cuts each streamline into short in-series fibres with their own junctions --- the
series-fibering that long muscles show \citep{roeleveld1997motor}; it is gated behind a
flag until validated.

\begin{figure}[tbp]
\centering
\figorbox[width=\textwidth]{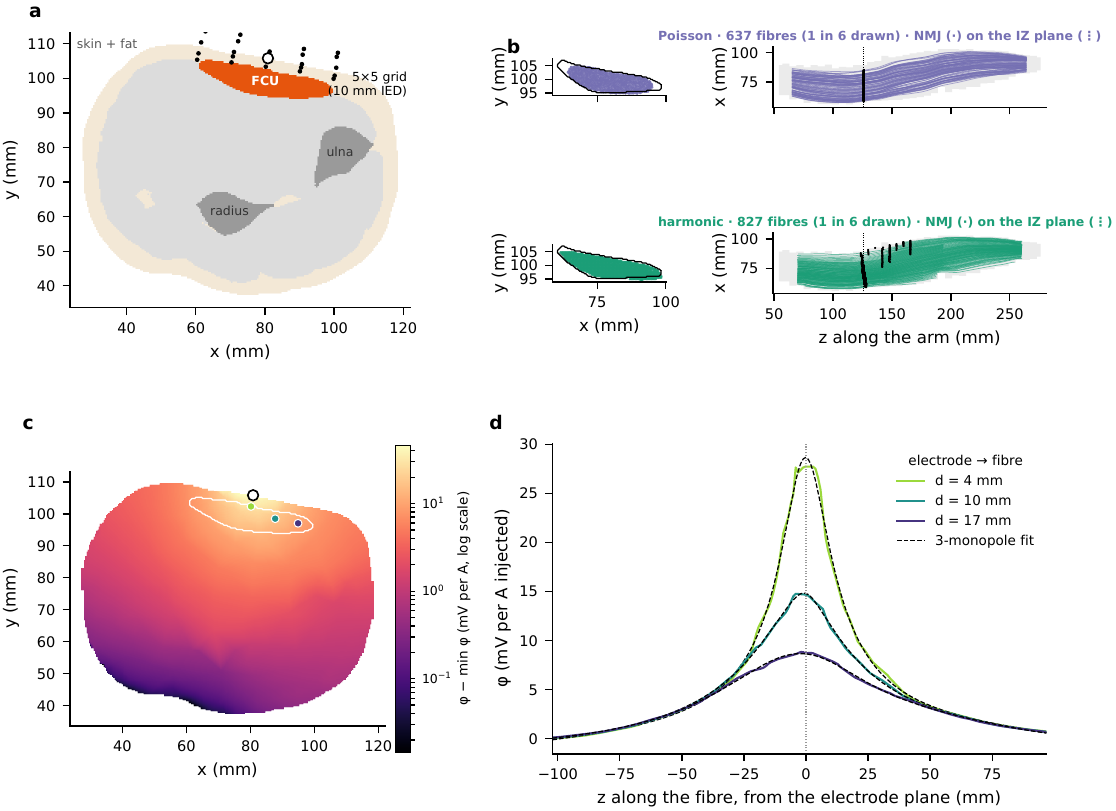}
\caption{The MRI forearm (subject WR). (a) A segmentation slice with the FCU highlighted
and the $5\times5$ electrode grid projected. (b) The FCU fibre bed at $4$~fibres/mm$^2$:
$637$ straight Poisson fibres and $827$ harmonic streamlines (one in six drawn), both
innervated on the same plane (dotted; the atlas fraction $0.305$ of the straight fibres'
length): $84\%$ of the streamlines cross it and are innervated within $1$~mm of it; the rest
start on its far side ($z = 140$--$165$~mm) and are innervated at their nearer end, the
second cluster of dots. (c) The reciprocal lead field of one
grid electrode on a mesh slice, $\varphi - \min\varphi$ on a log scale (as in
Fig.~\ref{fig:leadfield}d). (d) $\varphi$ along three fibres $4$, $10$ and $17$~mm from the
electrode, in mV per ampere injected, with the three-monopole fit the direct recipe uses
(dashed).}
\label{fig:mri}
\end{figure}

\emph{Why offer both.} On the FCU the two beds differ exactly where a straight model is
weakest: the $827$ streamlines (seeded at $0.5$~mm) stay inside the mask for $99.96\%$ of
their length against $96.6\%$ for the $637$ straight fibres, and their lengths follow
the muscle's shape rather than one global value (Fig.~\ref{fig:mri}b). Whether that
changes the signal is a question, not an assumption, and Sec.~\ref{sec:study-fibres}
answers it. One convention to watch: the harmonic model measures its junction fraction
along the muscle's principal axis, which for the FCU points the other way from the
straight bed's fibre coordinate; the pipeline therefore innervates both beds on the same
plane.

\subsection{Motor-unit pool}
\label{sec:pool}
\begin{figure}[tbp]
\centering
\figorbox[width=\textwidth]{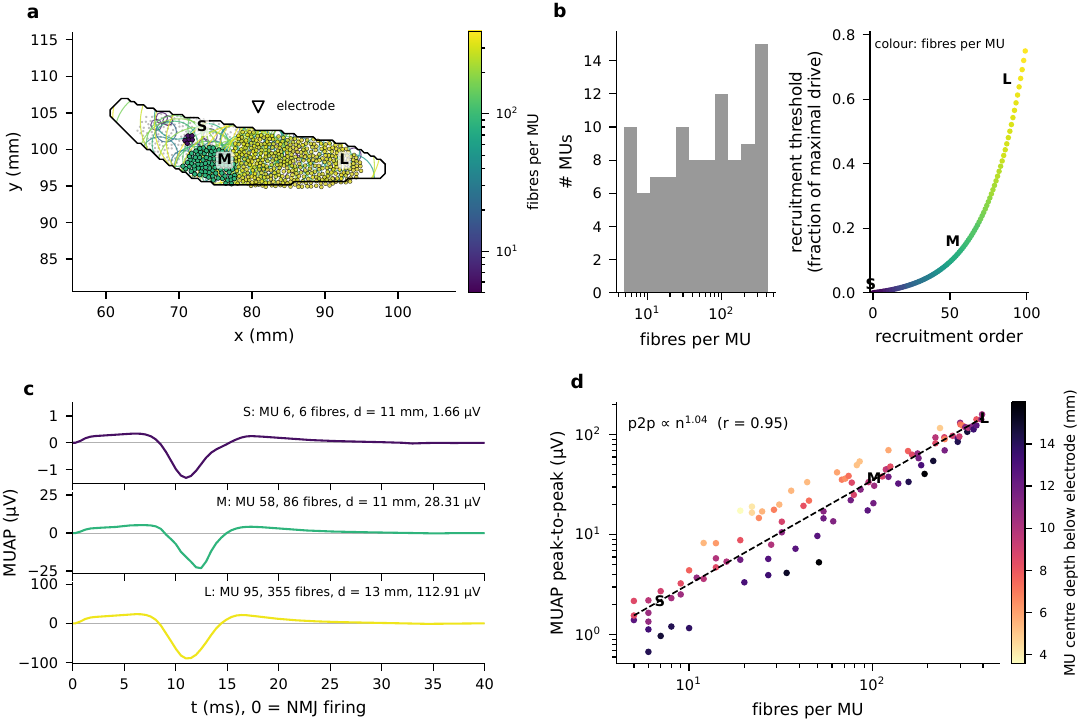}
\caption{The 100-unit FCU pool. (a) Territories on the cross-section, colour by size
($5$--$395$ fibres, median $66$; territory radii $0.6$--$13.8$~mm), with the centre
electrode of the grid. (b) Size distribution, and recruitment threshold against
recruitment order (colour: size). (c)
MUAPs at the centre electrode of a small, a medium and a large unit ($1.7$, $28$ and
$113$~$\mu$V; across the pool $0.7$--$160$~$\mu$V, median $24$, median duration $18$~ms).
(d) MUAP amplitude against fibre count: $p2p \propto n^{1.04}$ ($r = 0.95$), the scatter
being depth (colour); with shared fibres this is a consistency check (Sec.~\ref{sec:pool}).}
\label{fig:pool}
\end{figure}

Motor units are drawn from the bed following the size principle
\citep{fuglevand1993models}: $N$ sizes log-uniform between $5$ and $400$ fibres (many
small units, few large), each unit anchored on a random bed fibre and its territory
radius grown until it contains the target number of fibres, territories allowed to
overlap; units are sorted by size so that index $i$ is recruitment order. Because the bed
is a fixed sample of the muscle ($637$ fibres for $11{,}207$ fibre slots in the 100-unit
pool), fibres are shared between units --- $18$ units per fibre on average --- so a
fibre's SFAP is computed once and reused. Figure~\ref{fig:pool} shows the pool.

\emph{Why a shared bed.} The alternative, one private set of fibres per unit, costs a
lead-field sample and an SFAP per fibre slot ($17\times$ more) for no change in the
sum: two fibres $0.5$~mm apart have the same lead field to the precision of the mesh. The
price is that the amplitude--size relation in Fig.~\ref{fig:pool}d is a consistency
check (with shared fibres at similar depths, amplitude must scale with fibre count)
rather than evidence.

\subsection{From lead field to action potential: direct line-source synthesis}
\label{sec:direct}
We call the synthesis step \emph{direct line-source synthesis}: the line-source integral
is evaluated directly, in space, on the lead field sampled along the fibre. Its steps are
few and each has a reason.

\paragraph{The integral.} The action potential is the Rosenfalck form
$V(s) = A\,s^{3}e^{-s}$ for $s>0$ ($s$ in mm, $A = 96$~mV) \citep{rosenfalck1969intra},
launched from the neuromuscular junction at $z_{0}$ in both directions and cut at the
tendons by a window $w(z)$ over $[z_{0}-L_{1},\, z_{0}+L_{2}]$:
\begin{equation}
  \Vm(z,t) = V\!\big(v t - |z - z_{0}|\big), \qquad
  \Vm^{w}(z,t) = \Vm(z,t)\,w(z) ,
  \label{eq:wave}
\end{equation}
so that $t=0$ is the instant the junction fires and the wave front reaches the proximal
tendon at $L_{1}/v$ and the distal one at $L_{2}/v$. A fibre of radius $a$ and
intracellular conductivity $\sigma_{\mathrm{in}}$ injects, per unit length, the membrane
current of core-conductor theory \citep{rosenfalck1969intra, plonsey2007bioelectricity},
\begin{equation}
  i_{\mathrm m}(z,t) = \sigma_{\mathrm{in}}\,\pi a^{2}\,\frac{\partial^{2}\Vm^{w}}{\partial z^{2}} ,
  \label{eq:im}
\end{equation}
and the potential at the electrode is the lead-field-weighted sum of that current,
\begin{equation}
  \mathrm{SFAP}(t) = \int i_{\mathrm m}(z,t)\,\varphi(z)\,\dd z
  \;=\; \sigma_{\mathrm{in}}\,\pi a^{2}\int \Vm(z,t)\,w(z)\,\varphi''(z)\,\dd z ,
  \label{eq:sfap}
\end{equation}
where the second form follows by integrating by parts twice and shows that the SFAP is
the inner product of the windowed action potential with the \emph{second derivative} of
the lead field. Taking $\partial^{2}/\partial z^{2}$ of the whole windowed field, rather
than of each travelling half analytically, produces the three source terms that the
literature separates by hand \citep{dimitrov1998precise, kleinpenning1990equivalent}: the
two propagating tripoles, the \emph{generation} term at the cusp $|z-z_0|$, and the
\emph{end-of-fibre} terms where the window cuts the wave. These are exactly the terms
that make the source monopole-free, $\int i_{\mathrm m}\,\dd z = 0$ for all $t$
\citep{petersen2019comprehensive}; we verify this to $8\times10^{-17}$
(Sec.~\ref{sec:val-first}). $\varphi$ from the finite-element stage is the potential in
volts for $1$~A injected at the electrode; the engine uses $\sigma_{\mathrm{in}} = 1$~S/m
and $a = 50$~$\mu$m, so the constant $\sigma_{\mathrm{in}}\pi a^{2}$ is a scale factor
rather than a calibrated quantity and absolute microvolt values carry that caveat.

\begin{figure}[tbp]
\centering
\figorbox[width=\textwidth]{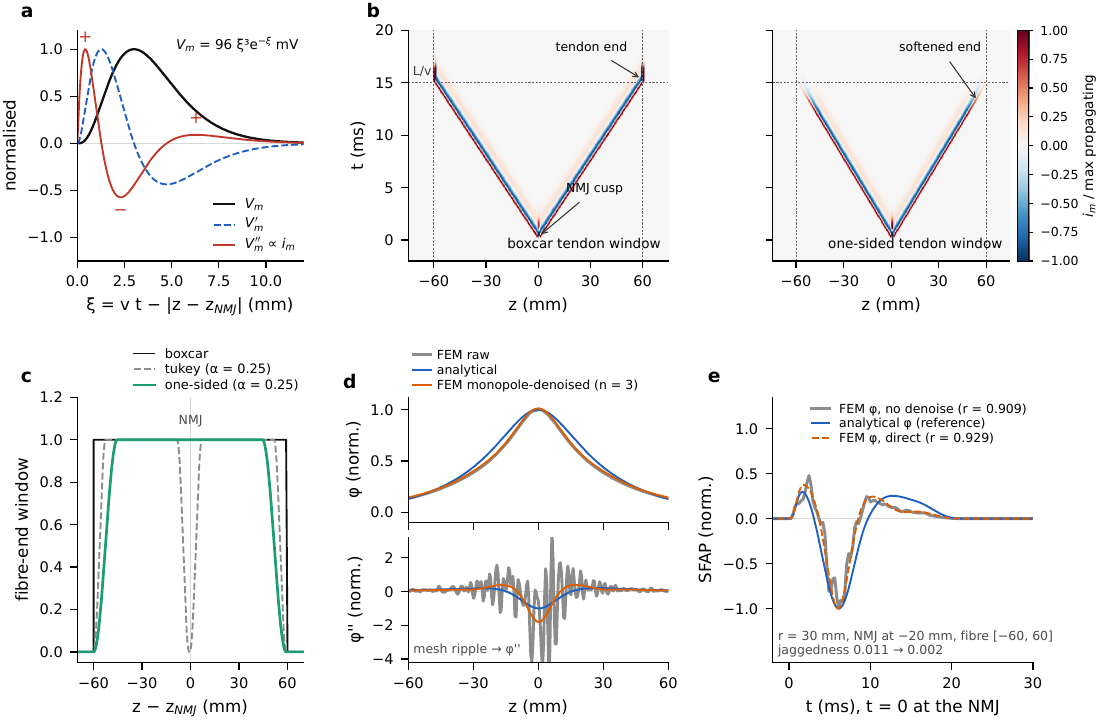}
\caption{Anatomy of direct line-source synthesis. (a) The Rosenfalck action potential and its
first and second spatial derivatives; the second derivative is the triphasic current
source. (b) The current-source-density matrix $\mathbf{C}(z,t)$ for a fibre
$[-60, 60]$~mm with the junction at $0$: two tripoles leave the junction, the cusp at
the junction and the window edges at the tendons appear as sources; boxcar (left) and
one-sided (right) tendon windows. (c) The three fibre-end windows. (d) A finite-element
lead field raw, monopole-denoised and analytical, with their second derivatives
(normalised to the analytical peak; the colour scale in (b) is clipped at the propagating
maximum --- the tendon term is $2.7\times$ larger). (e) The SFAPs each produces: the
ripple in $\varphi''$ reaches the raw SFAP and the monopole fit removes it ($r$ against the
analytical-$\varphi$ SFAP $0.909 \to 0.929$); the remaining difference from the analytical
waveform is the FEM field's own shape at this shallow depth (Sec.~\ref{sec:val-cylinder}),
not ripple.}
\label{fig:recipe}
\end{figure}

\paragraph{The recipe.} Discretised, Eq.~\eqref{eq:sfap} is one matrix--vector product,
$\mathrm{SFAP} = (\mathbf{C}\,\boldsymbol{\varphi})\,\Delta z$, with $\mathbf{C}$ the
$(n_t \times n_z)$ current-source-density matrix. The decisions around it, each with its
reason, are (Fig.~\ref{fig:recipe}, Table~\ref{tab:recipe}):
\begin{enumerate}
\item \textbf{Monopole denoise of $\varphi$.} A finite-element $\varphi(z)$ carries
  mesh-scale ripple, and the SFAP integrates $\varphi''$, which amplifies that ripple
  quadratically in spatial frequency. Instead of low-passing --- which rounds the real
  peak and halves the spectral content (Sec.~\ref{sec:val-features}) --- we replace
  $\varphi$ by a least-squares fit of the form the field actually has,
  $\hat\varphi(z) = \sum_{i=1}^{3} A_i/\sqrt{d_i^{2}+(z-z_i)^{2}} + c$, a sum of three
  free-position point sources. On a clean analytical $\varphi$ the fit is a near no-op
  ($r>0.99$); on a FEM $\varphi$ it changes the field by at most $1.8\%$ of its peak but
  lifts the correlation of $\varphi''$ with the analytical one from $0.47$ to $0.91$ and
  takes the SFAP's jaggedness from $0.011$ to $0.002$ (analytical: $0.0018$;
  Fig.~\ref{fig:recipe}d,e).
\item \textbf{Edge taper.} A short cosine ramp ($5$ and $10$ samples) on the two ends of
  the sampled $\varphi$, a Gibbs guard for lead fields truncated by the sampling window.
\item \textbf{Upsampling $\times 2$} (cubic spline). The second derivative needs a finer
  grid than the raw $\varphi$ or the SFAP develops a Nyquist zigzag; $\times2$ removes
  it and $\times4$ adds nothing.
\item \textbf{Current-source density.} $\mathbf{C}$ is built from the full bidirectional
  windowed field of Eq.~\eqref{eq:wave} by two numerical $z$-derivatives, so the
  generation and end-of-fibre terms come out of the same operation as the tripoles.
\item \textbf{One-sided tendon window.} $w(z)$ is flat through the junction and
  cosine-tapered over the last $25\%$ of each semi-fibre only. A symmetric window would
  notch the junction, where the wave is born; a hard boxcar produces a sharp end-of-fibre
  spike (right for one fibre with a sharp tendon, and the choice we use against
  sharp-tendon oracles) which in real muscle is softened by tendon scatter
  (Fig.~\ref{fig:phys}c). The taper costs $r=0.998$ against the sharp-tendon oracle.
\item \textbf{Physical time.} No centring: $t=0$ is the junction, the detector lobe of
  an electrode $\Delta z$ from the junction lands at $\Delta z/v$, the end-of-fibre
  onset at $L/v$ (Fig.~\ref{fig:phys}a); a $10$~ms pre-roll is prepended so a junction
  under the electrode is not pinned to the window edge. Because time is physical, junction
  scatter, conduction-velocity scatter and tendon scatter across the fibres of a unit
  produce the temporal dispersion real MUAPs show rather than a window artefact: a $5$~mm
  junction scatter across $50$ fibres lengthens a MUAP from $20.0$ to $22.2$~ms and
  reduces its peak-to-peak to $0.86$ of the coherent sum.
\end{enumerate}

\begin{table}[t]
\centering\small
\caption{The direct line-source recipe (\texttt{SpatialConfig}). The cylinder tier uses
$f_s = 4096$~Hz; the MRI tier $2048$~Hz with $v = 3.3$--$4$~m/s.}
\label{tab:recipe}
\begin{tabular}{llp{72mm}}
\toprule
parameter & value & why \\
\midrule
\texttt{denoise} & monopole, 3 poles & fit the analytic form of $\varphi$; removes mesh ripple before $\varphi''$ \\
\texttt{edge\_taper\_left/right} & 5 / 10 samples & Gibbs guard at the sampling-window ends \\
\texttt{upsample\_factor} & 2 & second derivative off the Nyquist zigzag \\
\texttt{csd\_derivative} & 2 & physical current source (triphasic) \\
\texttt{fiber\_window} & one-sided, $\alpha=0.25$ & flat at the junction, tapered at the tendons \\
\texttt{center\_time / t\_start\_ms} & false / $-10$ & physical time, junction at $t=0$ \\
$v$, $f_s$, $w$ & 4 m/s, 4096 Hz, 256 & coupled grid $\Delta z = v/f_s$ \\
$\sigma_{\mathrm{in}}$, $a$ & 1, 0.05 mm & amplitude constant $\sigma_{\mathrm{in}}\pi a^2$; no $1/v$ (the IAP is fixed in space) \\
\bottomrule
\end{tabular}
\end{table}

\paragraph{Why not the spatial-frequency route.} The pipeline also carries a
spatial-frequency-domain engine after \citet{farina2001novel, farina2004multilayer} ---
$\varphi$ transformed along $z$, multiplied by the fibre-end function, the
action-potential spectrum and $\mathrm{j}k_z$, brought back to time by a Radon section ---
and a line-by-line port of the Farina generator. It was our first engine and the
reference the direct method was originally tuned against. Given the same $\varphi$, our
port does not reproduce the line-source integral: it is anti-phase and lagged by
$0.9$--$2.6$~ms while its amplitude agrees within $5\%$ (Sec.~\ref{sec:val-first}), and in
the ported generator the end-of-fibre onset leads $L/v$ by $2.5$~ms and the propagating
main lobe has the opposite sign to the textbook convention. We attribute these to the
orientation and sign conventions of the action potential in our port --- the original
model was verified against limiting cases and reproduced by finite elements to $3$--$5\%$
\citep{maksymenko2023myoelectric} --- and until that audit is done the direct method
carries production and the other two are kept for comparison.

\paragraph{From SFAPs to a MUAP.} A \emph{fibre bed} object carries, per fibre, the
sampling step (arc length for curved fibres), the two semi-lengths, the junction position
and the conduction velocity; a MUAP is the sum of the SFAPs of the fibres in a motor
unit, each computed from its own $\varphi(z)$ with the shared configuration, on every
electrode of the layout.

\subsection{Activation: from motor units to signals}
\label{sec:activation}
The activation layer is phenomenological, written from the equations of the Fuglevand
pool as parametrised in NeuroMotion \citep{fuglevand1993models, ma2024neuromotion}. Given
a scalar excitation $E(t)\in[0,1]$: unit $i$ has a recruitment threshold spread
exponentially with range $50$ up to $E=0.75$ for the last unit; once recruited its rate is
$\min\big(\mathrm{PFR}_i,\ \mathrm{MFR}_i + (E-\mathrm{RTE}_i)\,s_i\big)$ with the peak
rate falling from $40$ to $30$~pps, the minimum from $10$ to $5$~pps and the slope from
$50$ to $30$~pps per unit drive across the pool --- the onion skin
\citep{deluca2010relationship}; spikes are a renewal process with Gaussian inter-spike
intervals of coefficient of variation $1/6$, re-drawn at each discharge, and a common-drive
fluctuation (low-passed noise at $2$~Hz) can be added to $E$. Each discharge adds a
twitch $P_i (t/T_i)\,e^{1-t/T_i}$ with amplitudes spanning $100\times$ and larger units
contracting faster, scaled by the rate-dependent fusion gain of
\citet{fuglevand1993models}; the sum is normalised to the mean force at full excitation,
so force is in \%MVC. Each spike places the unit's MUAP (single channel or on every
electrode) and the trains sum; single- and double-differential montages are differences
of electrodes. For \emph{movement}, a normalised joint angle $\theta(t)$ sets the drive
($0.08 + 0.42\,\theta$), scales each spike's MUAP ($1 + g_a(\theta-0.5)$; shortening brings
fibres to the electrode) and time-warps it ($1 + g_w(\theta-0.5)$), with defaults
$g_a = 0.7$, $g_w = 0.18$.

\begin{figure}[tbp]
\centering
\figorbox[width=\textwidth]{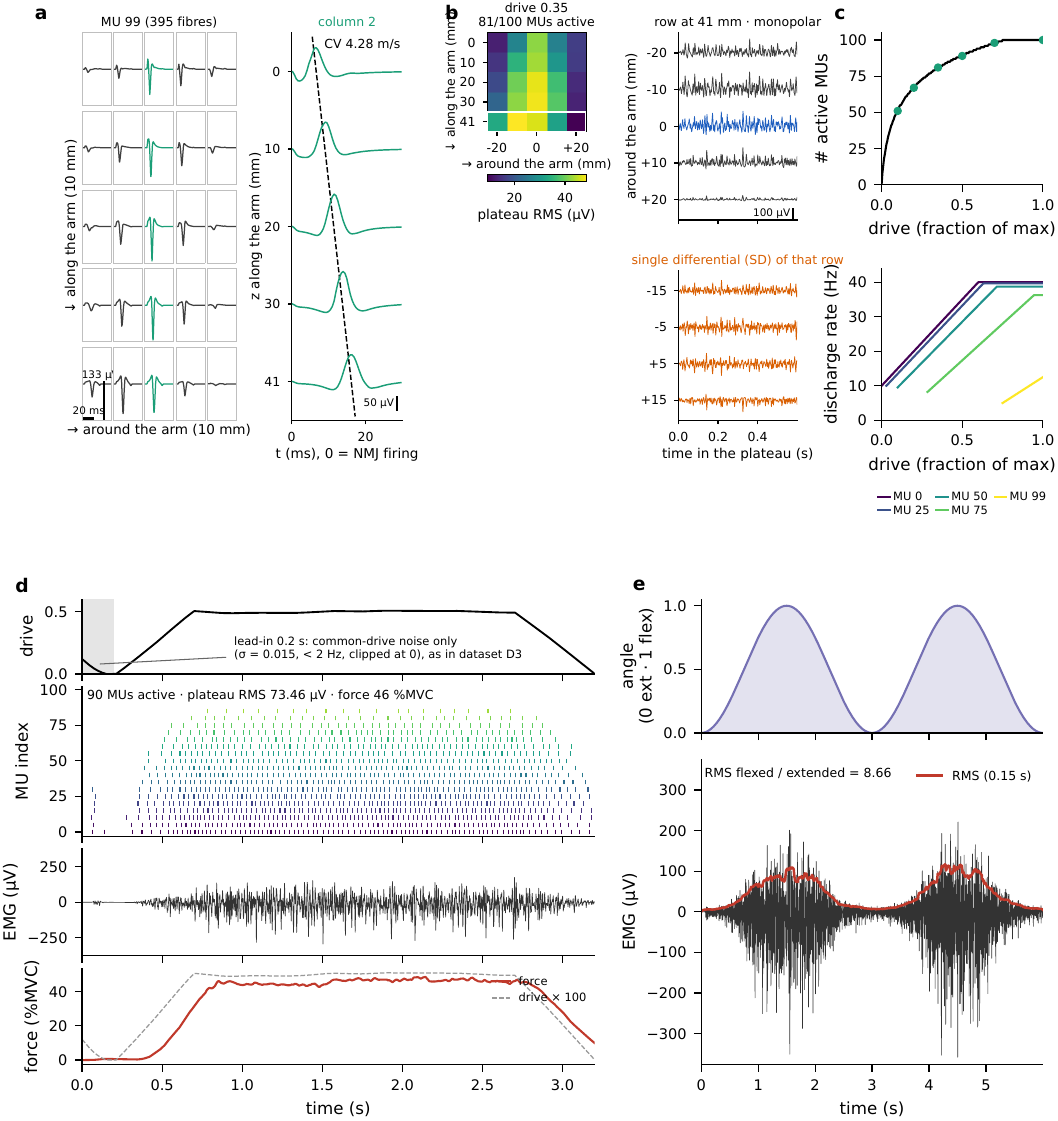}
\caption{From motor units to signals (regular $5\times5$ grid, $10$~mm spacing). (a) The
largest unit's MUAP on the grid --- rows run along the arm, columns around it --- and,
right, the column with the largest amplitude as a waterfall offset by electrode position:
the wave walks down the column at $4.28$~m/s (dashed guide; $4.1$--$4.8$ over the pool,
set $4.0$). (b) Interference HD-EMG at a $35\%$-drive plateau ($81$ of $100$ units
active): the RMS map of the grid ($9$--$48$~$\mu$V), one row of monopolar traces and the
single-differential traces of that row at a shared scale. (c) Recruitment and rate coding of the pool. (d) A trapezoid contraction at
half drive: drive (the lead-in carries the common-drive fluctuation), spikes, EMG (plateau
RMS $73$~$\mu$V), force ($46\%$~MVC); across drives $0.1$--$1.0$ the plateau RMS is
$14$--$133$~$\mu$V and the force $7$--$100\%$~MVC. (e) A movement trial: angle and
non-stationary EMG (RMS flexed/extended $8.7$).}
\label{fig:hdemg}
\end{figure}

\emph{Why phenomenological.} The pool is the best-tested part of the EMG literature and
its parameters are published; a biophysical motoneuron model would add cost and free
parameters without a validation target at the surface. The movement layer is a modulation,
not a musculoskeletal model; models of the geometry change itself exist
\citep{mesin2006finite, mesin2011insights} and are the next step. Figure~\ref{fig:hdemg}
shows a unit on the grid, an interference plateau, the pool's recruitment and rate coding,
a trapezoid contraction with its force, and a movement trial.

\section{Validation}
\label{sec:validation}

Every stage above can be tested in isolation, and the suite that ships with the code
(\texttt{scripts/validation/}, $\approx4$~min) does so: fifty checks, each holding one
measurable property against a literature principle with a numeric criterion, in four
tiers --- (A) the cylinder must reproduce: first principles, then the analytical model,
then the finite-element model, then the pipeline on each; (B) the MUAP phenomenology the
literature agrees on; (C) the statistics of interference EMG and the pool; (S) chain-level
sanity checks. Table~\ref{tab:score} is the scoreboard and Appendix~\ref{app:checks}
lists every check with its criterion and measured value. A ``known'' flag marks a
documented limitation that keeps its criterion, so it turns green when fixed without
failing the run. This section gives the results that matter for a reader of the pipeline;
the open items are collected in Sec.~\ref{sec:discussion}.

\begin{table}[t]
\centering\small
\caption{Validation scoreboard (branch \texttt{paper/arxiv-v1}; tiers C and S run on the
released forearm pool, dataset D2). ``pass / total'' counts every check; ``known'' are
documented limitations that fail without failing the tier; the two genuine failures are
discussed in Sec.~\ref{sec:discussion}. Appendix~\ref{app:checks} lists all $50$ checks.}
\label{tab:score}
\begin{tabular}{lp{58mm}ccp{42mm}}
\toprule
tier & scope & pass / total & known & fail \\
\midrule
A & cylinder: first principles, analytical, FEM, pipeline & 16 / 20 & 3 & A1.1b: FEM $\varphi''$ at mid-depth, $r=0.93$ \\
B & MUAP features vs literature & 14 / 14 & 0 & --- \\
C & interference EMG and pool & 6 / 8 & 1 & C2: no refractory floor \\
S & chain-level sanity & 8 / 8 & 0 & --- \\
\bottomrule
\end{tabular}
\end{table}

\subsection{The synthesis step against first principles}
\label{sec:val-first}
\begin{figure}[tbp]
\centering
\figorbox[width=\textwidth]{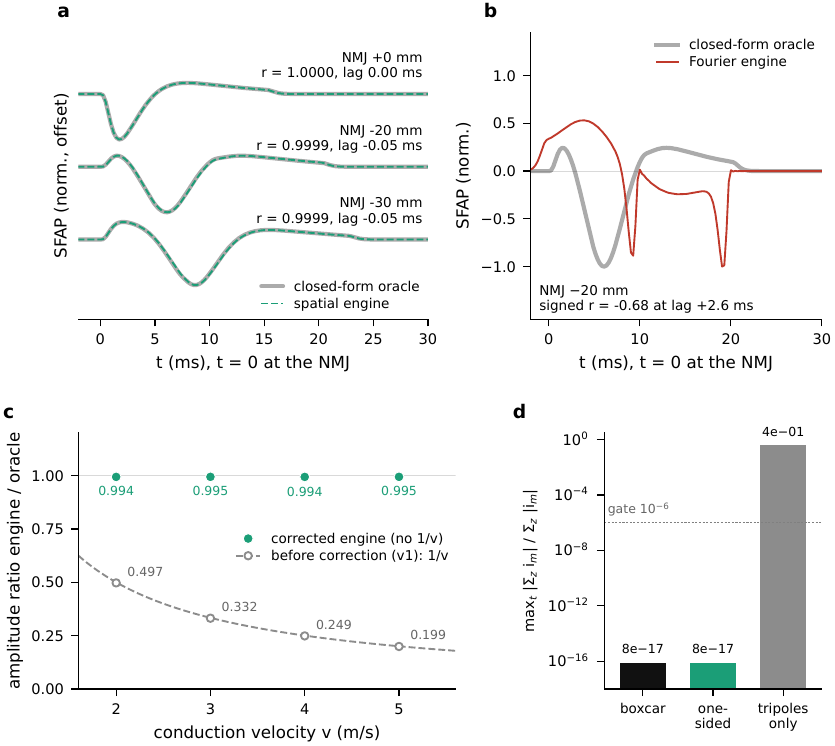}
\caption{First principles. (a) The closed-form line-source SFAP and direct line-source
synthesis, three junction positions. (b) The spatial-frequency engine against the same
oracle. (c) The engine-to-oracle amplitude ratio: $0.994$--$0.995$ at every $v$ after
the correction described in the text; the dashed $1/v$ is what the uncorrected engine
gave. (d) The net source current at every instant is zero to machine precision once the
junction and end terms are included; the propagating tripoles alone leave $39\%$.}
\label{fig:first}
\end{figure}

A fibre in an infinite anisotropic medium has a closed-form lead field,
$\varphi(z) \propto (\rho^{2}/\sigma_\rho + z^{2}/\sigma_z)^{-1/2}$
\citep{plonsey2007bioelectricity, rush1963resistivity}, whose second derivative is
analytic. Evaluating Eq.~\eqref{eq:sfap} in the dual form on a $0.02$~mm grid gives an
oracle with no numerical derivative and no volume-conductor model in the loop. Direct
line-source synthesis, given the same $\varphi$ on its $0.977$~mm grid, matches it with
$r = 1.0000$, $0.9999$, $0.9999$ and lags $0.00$, $-0.05$, $-0.05$~ms for junctions $0$,
$-20$, $-30$~mm from the electrode (Fig.~\ref{fig:first}a). The source is monopole-free to
$7.7\times10^{-17}$; the SFAP is invariant to the sampling rate and step ($r>0.998$,
amplitude within $0.3\%$); moving the electrode by $20$~mm delays the propagating lobe by
$4.88$~ms ($\Delta z/v = 5.00$, one sample); superposition is exact; and changing $v$
from $4$ to $3$~m/s stretches the SFAP by $4/3$ and scales its mean frequency by $3/4$
\citep{lindstrom1977interpretation}. The oracle also caught a defect: the first version
of the engine divided by $v$ once too often (a constant inherited from the
spatial-frequency formulation, where the $1/v$ belongs), so its ratio to the oracle was
exactly $1/v$. With the division removed the ratio is $0.994$ at every $v$ from $2$ to
$5$~m/s (Fig.~\ref{fig:first}c). The spatial-frequency engine on the same $\varphi$
scores $r = -0.79$, $-0.68$, $-0.64$ at lags of $1.4$, $2.6$, $0.9$~ms
(Fig.~\ref{fig:first}b).

\subsection{The finite-element conductor against the analytical cylinder}
\label{sec:val-cylinder}
Along fibres $7$--$20$~mm below the skin the FEM $\varphi(z)$ correlates with the
analytical one at $r = 0.995$--$1.000$, and at $r=0.997$--$1.000$ for fibres
$10$--$45^\circ$ off the electrode meridian. Their widths differ for the shallowest
fibres (FWHM $41$ vs $35$~mm at $7$~mm), where the FEM Gaussian source behaves as a
$5$~mm-diameter disc; at mid-depth ($13$~mm) $\varphi''$ agrees only to $r=0.93$ whatever
electrode is assumed --- the one tier-A failure. The pipeline run on FEM $\varphi$ against
the pipeline run on analytical $\varphi$ gives $r = 0.90$--$0.97$ for shallow fibres and
$0.998$ for deep ones. Timing and symmetries hold: the end-of-fibre onset lands at
$10.26$ and $15.39$~ms for $L/v = 10$ and $15$; conduction velocity from a four-electrode
array is $3.98$~m/s for $v=4$; rotating the electrode by $30^\circ$ equals rotating the
fibre ($r=1.000$); shifting it $20$~mm along the axis equals shifting the sampling window
($r=0.996$--$0.999$, the residual being the $240$~mm mesh's end effect
\citep{lowery2002multiple}); and the lead field between two interior points is
reciprocal to $1.35\%$. Two finite-element items stay open (Sec.~\ref{sec:discussion}):
on FEM $\varphi$ the SFAP amplitude is erratic at $\pm40\%$ across depth while shape,
timing and spectrum are faithful, and the FEM cylinder decays $\approx30\%$ slower with
depth than the analytical one.

\subsection{MUAP phenomenology}
\label{sec:val-features}
\begin{figure}[tbp]
\centering
\figorbox[width=\textwidth]{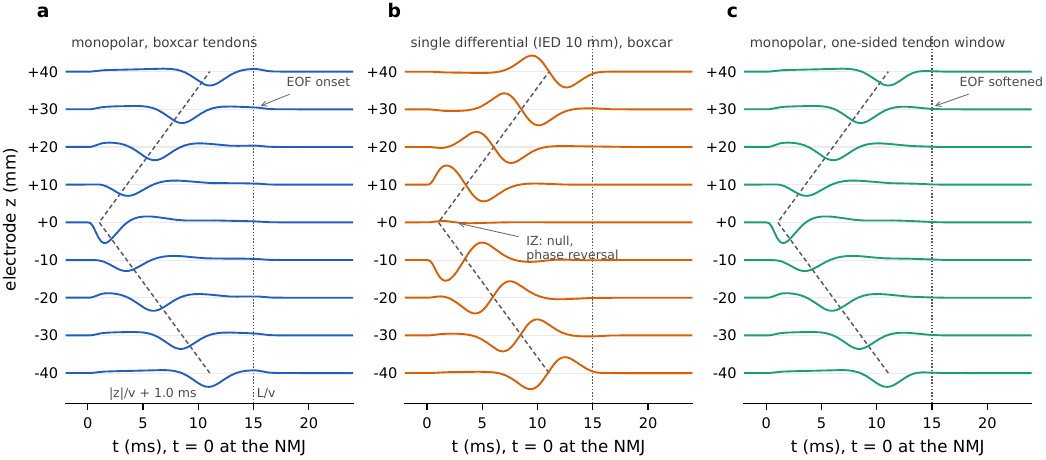}
\caption{Physical time on a finite-element lead field (fibre 10~mm deep, tendons at
$\pm60$~mm, $v = 4$~m/s). (a) Monopolar SFAPs along a nine-electrode array with hard
tendons: the propagating lobe walks at $|z|/v$ (a fit gives $4.04$~m/s; the guide is drawn
at $|z|/v + 1.0$~ms because the negative phase trails the wave front by the action
potential's rise) while the end-of-fibre potential starts at $15.4$--$15.6$~ms on every
channel ($L/v = 15$~ms; spread $0.24$~ms). (b) The single-differential montage vanishes and reverses at the innervation
zone. (c) The same array with the one-sided tendon window of the direct recipe: the
end-of-fibre potential falls from $0.16$ to $0.10$ of the propagating amplitude.}
\label{fig:phys}
\end{figure}

Each check is run on our pipeline and on the ported Farina generator, with arrays and
montages built by translating the lead field (Figs.~\ref{fig:phys}, \ref{fig:features}).
Between the junction and the tendon the monopolar SFAP is $+\,-\,+$ with the main lobe
negative \citep{merletti2019tutorial}. Conduction velocity from a single-differential
array recovers $3.13$, $3.98$, $4.88$~m/s for set values $3$, $4$, $5$
\citep{farina2004estimation}. Monopolar potentials mirror about the junction ($r>0.998$);
the single-differential channel over it is $4\%$ of the maximum and the channels either
side are mirror images of opposite sign \citep{masuda1985position}. The end-of-fibre
potential has the same onset on every electrode; its ratio to the propagating part grows
monotonically with depth and is suppressed mono $>$ SD $>$ DD
\citep{arabadzhiev2013peculiarities, roeleveld1998motor}. Amplitude is a power law in
depth over $7$--$25$~mm ($R^{2}>0.98$) \citep{roeleveld1997motor,
fuglevand1992detection}; RMS falls with subcutaneous fat as in the finite-element curve of
\citet{kuiken2003effect}; amplitude and mean frequency fall with electrode radius
\citep{merletti2019tutorial}; with a single travelling wave the two electrodes of a
bipolar pair see shift-copies to $2\times10^{-3}$, so the montage is the textbook comb
filter with nulls at $n\,v/\mathrm{IED}$ \citep{lindstrom1977interpretation,
lynn1978influence}; anisotropy elongates the lead field along the fibres by $\sqrt{5}$ in
the infinite medium and by $1.5$ in the layered cylinder; and $50$ identical fibres give
$50\times$ the SFAP to $10^{-15}$. The numbers behind these are the material of
Sec.~\ref{sec:study-electrode}.

\subsection{Interference EMG and the pool}
\label{sec:val-emg}
On the released forearm pool the rates lie in $5$--$40$~pps with the onion-skin ordering
($\rho = -0.98$ between threshold and rate); inter-spike-interval variability is $0.16$;
thresholds are right-skewed and twitches span $95\times$; the amplitude distribution is
super-Gaussian at low drive (kurtosis $5.6$ at $5\%$) and Gaussian at high ($2.9$--$3.1$
at $50$--$90\%$), with ARV/RMS $0.77$--$0.79$, between the Laplacian and Gaussian values
\citep{clancy1999probability, nazarpour2013note}; the RMS at half maximal force is $0.56$
of that at maximal force, inside the linear-to-quadratic band of
\citet{lawrence1983myoelectric}; larger units give larger MUAPs ($\rho = 0.98$). The
chain-level checks pass: orderly recruitment, onion skin, EMG rising with drive, force
monotone and calibrated to $100\%$~MVC at full drive, a localised MUAP footprint, an
$8.7\times$ amplitude modulation across a movement, and a MUAP scale inside the
physiological band (median peak-to-peak $24$~$\mu$V, range $0.7$--$160$~$\mu$V, median
duration $18$~ms, against tens to hundreds of microvolts and $5$--$20$~ms
\citep{merletti2019tutorial}). Conduction velocity read off the HD grid is $4.33$~m/s
($4.1$--$4.9$ over the $100$ units) for a set $4.0$~m/s: the $8$--$24\%$ overestimate is
the apparent velocity $v/\cos\theta$ of fibres running $5$--$25^\circ$ to the grid columns
(the same estimator recovers $3.98$ on the cylinder; inclination bias of this size is
documented \citep{mesin2007estimation}). The Gaussian renewal process has no refractory
floor ($1.6\%$ of intervals below $20$~ms at $40$~pps) --- the one tier-C failure.

\section{Studies with the pipeline}
\label{sec:studies}

The pipeline is a tool; this section uses it. Each study asks one question a user of
surface EMG might ask, answers it with the pipeline, and ends with what to take from it.

\subsection{What the electrode sees: depth, fat, spacing, montage}
\label{sec:study-electrode}
\begin{figure}[tbp]
\centering
\figorbox[width=\textwidth]{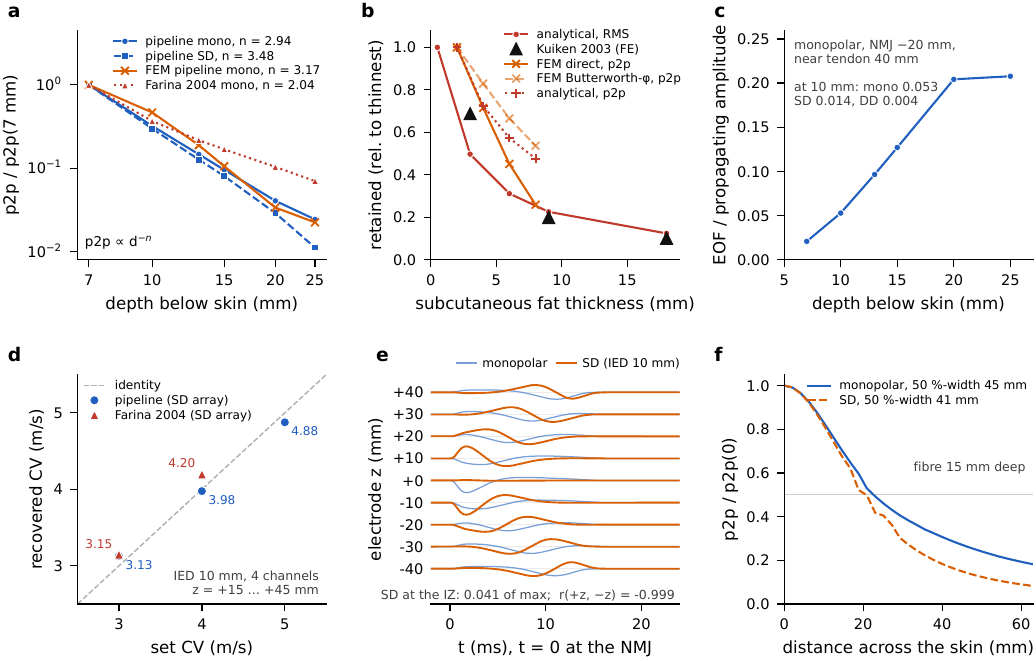}
\caption{MUAP features against the literature (fibre $10$~mm below the skin unless
stated; $v = 4$~m/s). (a) Depth law of the peak-to-peak amplitude (log--log) for the
pipeline (monopolar and single-differential), the Farina generator and the FEM pipeline,
with fitted exponents. (b) RMS retained against subcutaneous fat thickness, analytical
model against the finite-element points of \citet{kuiken2003effect}; FEM pipeline
(direct and Butterworth-smoothed) and Farina generator at $2$--$8$~mm, each relative to
its thinnest case. (c) End-of-fibre to propagating amplitude ratio against depth
(junction $20$~mm and near tendon $40$~mm from the electrode). (d) Conduction velocity
recovered from a four-channel single-differential array against the set value.
(e) The monopolar and single-differential array across the innervation zone (IED
$10$~mm). (f) Transverse spread of monopolar and single-differential amplitude for a
fibre $15$~mm deep.}
\label{fig:features}
\end{figure}

On the validation cylinder ($v = 4$~m/s; fibre $10$~mm deep unless stated) the
peak-to-peak amplitude of a fibre falls with depth as a power law with exponent $2.9$
monopolar and $3.5$ single-differential over $7$--$25$~mm (Fig.~\ref{fig:features}a): a
fibre at $20$~mm gives $\approx8\%$ of the monopolar amplitude of one at $7$~mm and
$\approx3\%$ of the single-differential one. Subcutaneous fat attenuates the RMS to
$0.50$, $0.23$ and $0.12$ at $3$, $9$ and $18$~mm (Fig.~\ref{fig:features}b) and lowers
the mean frequency. The end-of-fibre potential --- the non-propagating component every
electrode sees at the same instant --- grows from $0.02$ of the propagating amplitude at
$7$~mm to $0.20$ at $20$~mm (Fig.~\ref{fig:features}c), and the montage suppresses it:
mono $0.053$, single-differential $0.014$, double-differential $0.004$ at $10$~mm. A
larger electrode low-passes the signal ($-2.8$~dB at $100$~cycles/m for a $5$~mm disc);
the single-differential amplitude doubles from an inter-electrode distance of $2.5$ to
$5$~mm and saturates beyond $10$~mm, and the bipolar montage is a comb filter with nulls
at $n\,v/\mathrm{IED}$. Across the skin, the amplitude of a fibre $15$~mm deep falls to
half at $45$~mm monopolar and $41$~mm single-differential from the fibre
(Fig.~\ref{fig:features}f).

\emph{What to take from it.} Surface EMG is a shallow measurement: the top $10$~mm of a
muscle dominate every montage, and units below $\approx20$~mm are represented mostly by
their end-of-fibre potential, which carries no conduction-velocity information and is
what differential montages remove. For selectivity, double-differential at
$5$--$10$~mm spacing removes most of the far-field component at the cost of amplitude;
for amplitude, monopolar or single-differential at $\geq10$~mm spacing. Fat thickness
should be reported with any amplitude comparison across subjects: $9$~mm of fat costs
$77\%$ of the RMS.

\subsection{Galleries: one parameter at a time}
\label{sec:study-waterfalls}
\begin{figure}[tbp]
\centering
\figorbox[width=\textwidth]{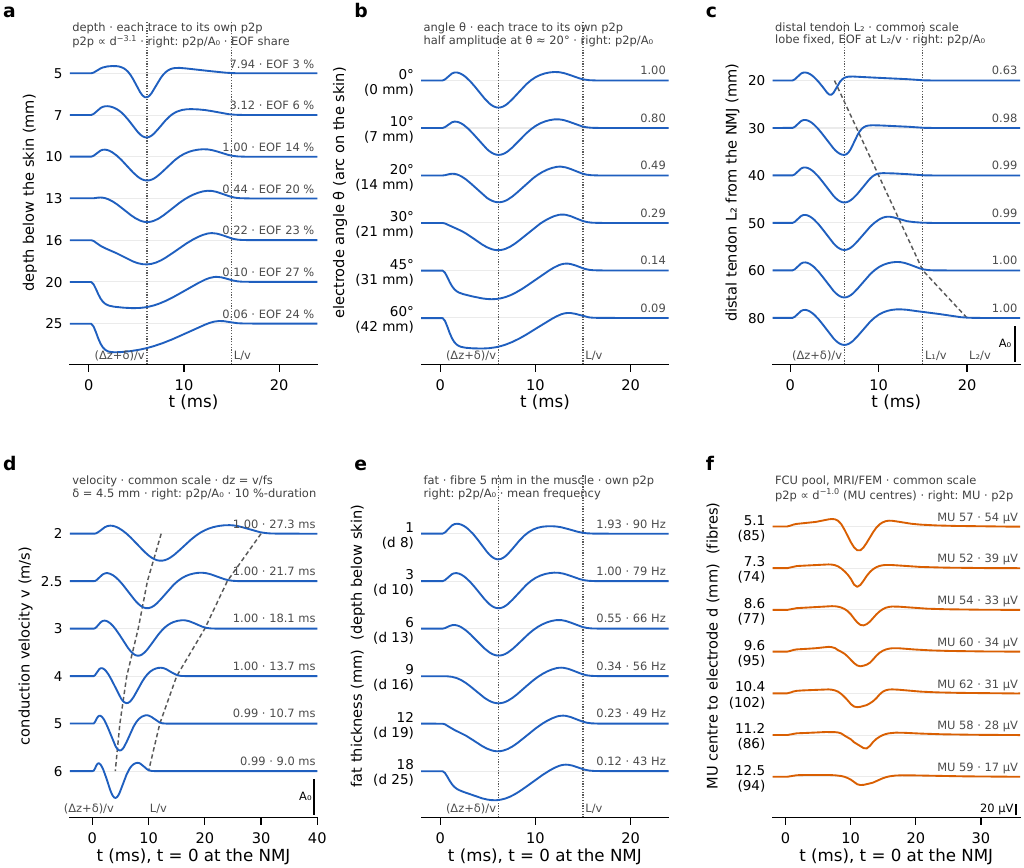}
\caption{Waterfall galleries, one parameter at a time (direct line-source recipe; $t=0$
at the junction; dotted guides: lobe arrival $(\Delta z+\delta)/v$ with $\delta = 4.5$~mm,
end-of-fibre at $L/v$). (a--e) Analytical cylinder, fibre $10$~mm deep, junction $20$~mm
from the electrode, tendons at $\pm60$~mm unless swept; amplitudes relative to that
reference fibre ($A_0$), peak-to-peak ratio on the right. (a) Depth $5$--$25$~mm, each
trace to its own amplitude, with the end-of-fibre share. (b) Electrode angle around the
cylinder $0$--$60^\circ$. (c) Distal tendon $20$--$80$~mm from the junction, common
scale; the end-of-fibre potential walks along $L_2/v$ (dashed). (d) Conduction velocity
$2$--$6$~m/s with the coupled grid $\Delta z = v/f_s$, common scale, with the $10\%$
duration. (e) Fat $1$--$18$~mm with the fibre $5$~mm inside the muscle (depth below the
skin in brackets), with the mean frequency. (f) MRI forearm: seven units of $74$--$102$
fibres at increasing distance from the centre electrode (unit centre to electrode; fibres
in brackets), common scale in $\mu$V.}
\label{fig:waterfalls}
\end{figure}

Figure~\ref{fig:waterfalls} shows the same potential as one parameter moves and nothing
else does --- on the analytical cylinder with the direct recipe (a--e; amplitudes relative
to the reference fibre, $10$~mm deep under $3$~mm of fat) and on the MRI forearm (f).
Depth is the strongest lever: from $5$ to $25$~mm the amplitude falls $142\times$
($\propto d^{-3.1}$), the negative lobe widens from $2.2$ to $8.5$~ms and the end-of-fibre
share grows from $3$ to $24\%$, while the wave's arrival at $(\Delta z+\delta)/v$ does not
move. Moving the electrode around the limb does the same in a milder form: half
amplitude at $20^\circ$ ($14$~mm of skin), a tenth at $60^\circ$, the timing unchanged.
Moving the distal tendon leaves the propagating lobe where it is and walks the
end-of-fibre potential along $L_2/v$, from $6$ to $20$~ms. Conduction velocity moves
everything at once: the lobe at $(\Delta z+\delta)/v$, the end-of-fibre at $L/v$, the
duration as $1/v$ ($27$ to $9$~ms from $2$ to $6$~m/s) and the mean frequency as $v$
($39$ to $115$~Hz), at constant amplitude. Fat, with the fibre held $5$~mm inside the
muscle, costs $16\times$ in amplitude and halves the mean frequency ($90$ to $43$~Hz)
from $1$ to $18$~mm. On the forearm, units of $74$--$102$ fibres at $5$ to $12.5$~mm from
the centre electrode fall from $54$ to $17$~$\mu$V, the lobe lagging further behind the
common junction-to-electrode delay as depth grows. These galleries are what a
decomposition algorithm or a decoder sees when one physiological quantity varies; they
come from one script and can be redrawn for any parameter of the chain.

\subsection{Does fibre geometry change the signal?}
\label{sec:study-fibres}
\begin{figure}[tbp]
\centering
\figorbox[width=0.86\textwidth]{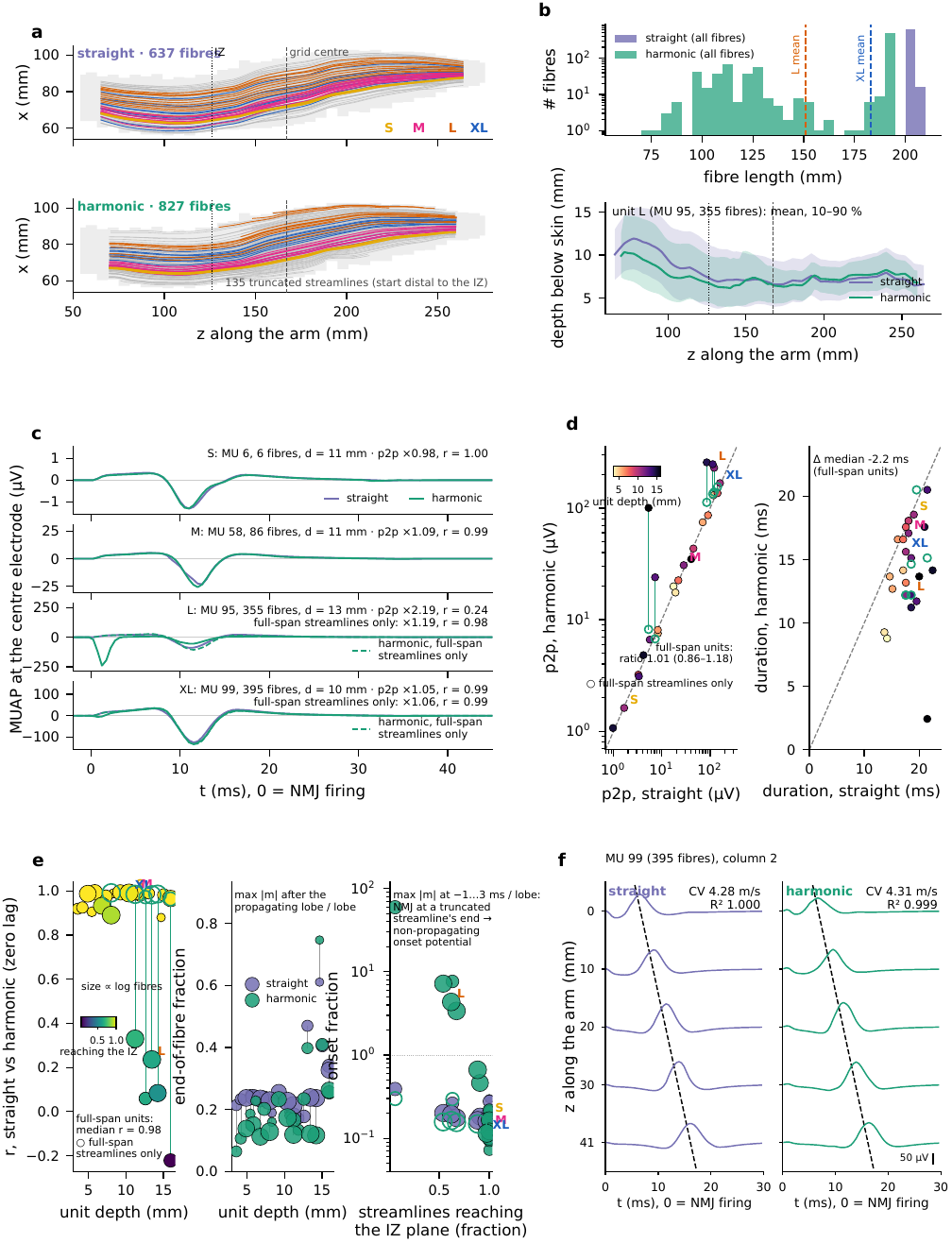}
\caption{Straight against harmonic-streamline fibre beds in the FCU: same units,
innervation plane, grid and recipe. (a) Side view of the two beds with the S/M/L/XL
units highlighted; $135$ of the $827$ streamlines start distal to the innervation plane.
(b) Fibre lengths (streamlines bimodal; straight fibres one length) and depth below the
skin along the fibres of unit L (mean and 10--90\% band): both beds vary in depth along
the fibre. (c) MUAPs of the four units at the centre electrode, straight and harmonic;
for L and XL also the harmonic unit re-drawn from full-span streamlines. (d) Peak-to-peak
amplitude and duration, harmonic against straight, colour by depth; open circles are the
full-span re-draw. (e) Waveform correlation, end-of-fibre fraction and onset fraction
(largest excursion in the first $3$~ms relative to the propagating lobe) against unit
depth or the fraction of the unit's streamlines that reach the innervation plane: the
units off the diagonal are the truncated-streamline cases. (f) Column waterfall of the
largest unit on both beds with the conduction velocity read off the grid.}
\label{fig:fibregeom}
\end{figure}

We re-drew $24$ units of the released FCU pool ($5$--$395$ fibres, $4$--$16$~mm deep,
including the three units of Fig.~\ref{fig:pool} and the largest unit) on the
harmonic-streamline bed with the same territory centres and sizes, the same innervation
plane and the same grid, and synthesised their MUAPs with the same recipe on the column of
largest amplitude (Fig.~\ref{fig:fibregeom}). For the $18$ units whose streamlines span
the muscle, geometry is a second-order effect: peak-to-peak amplitude changes by a median
factor of $1.01$ ($0.86$--$1.18$) with no dependence on depth or size, the waveforms
correlate at $r = 0.98$ ($0.88$--$1.00$) with lags below $0.5$~ms, the harmonic MUAPs
are $\approx2$~ms shorter with a $\approx20\%$ smaller end-of-fibre component (curved
paths reach the tendons at spread-out times), and the conduction velocity read off the
column agrees to $3\%$ ($4.28$ against $4.30$~m/s). The first-order effect is not
geometry but innervation. A sixth of the FCU streamlines are truncated --- they start
distal to the innervation plane, on the deep side of the muscle --- and the
single-junction rule places their junction at the fibre's end; the six units that draw on
them acquire a synchronous, non-propagating onset potential $0.7$--$60\times$ the
propagating lobe, a $2$--$19\times$ inflated amplitude, $r\approx0$ and no usable
conduction velocity (unit L in Fig.~\ref{fig:fibregeom}c). Re-drawing those units from
full-span streamlines removes the artefact ($r \geq 0.96$). At the fibre level the beds
differ where expected: streamline lengths are bimodal (median $193$~mm, a fifth below
$120$~mm) against one global $204$~mm, and containment is $99.9\%$ against $96.6\%$;
but --- a correction to the usual description --- the straight bed is not constant-depth
either, since its paths follow the muscle centreline (depth excursion $6.4$~mm along a
straight fibre, $5.2$~mm along a streamline).

\emph{What to take from it.} For HD-EMG synthesis the straight bed is an adequate
default: the same amplitudes, shapes and velocities within a few percent. Use the
harmonic bed when anatomy matters --- fibres that stay inside the muscle and follow its
shape --- and prune or re-innervate truncated streamlines first, because a junction at a
fibre's end is a far-field generator that no montage removes. Neither bed narrows the
surface field to real motor-unit widths; that is the job of the series-fibering variant,
not of curvature.

\subsection{Crosstalk: what a grid over one muscle receives from its neighbours}
\label{sec:study-crosstalk}
\begin{figure}[tbp]
\centering
\figorbox[width=\textwidth]{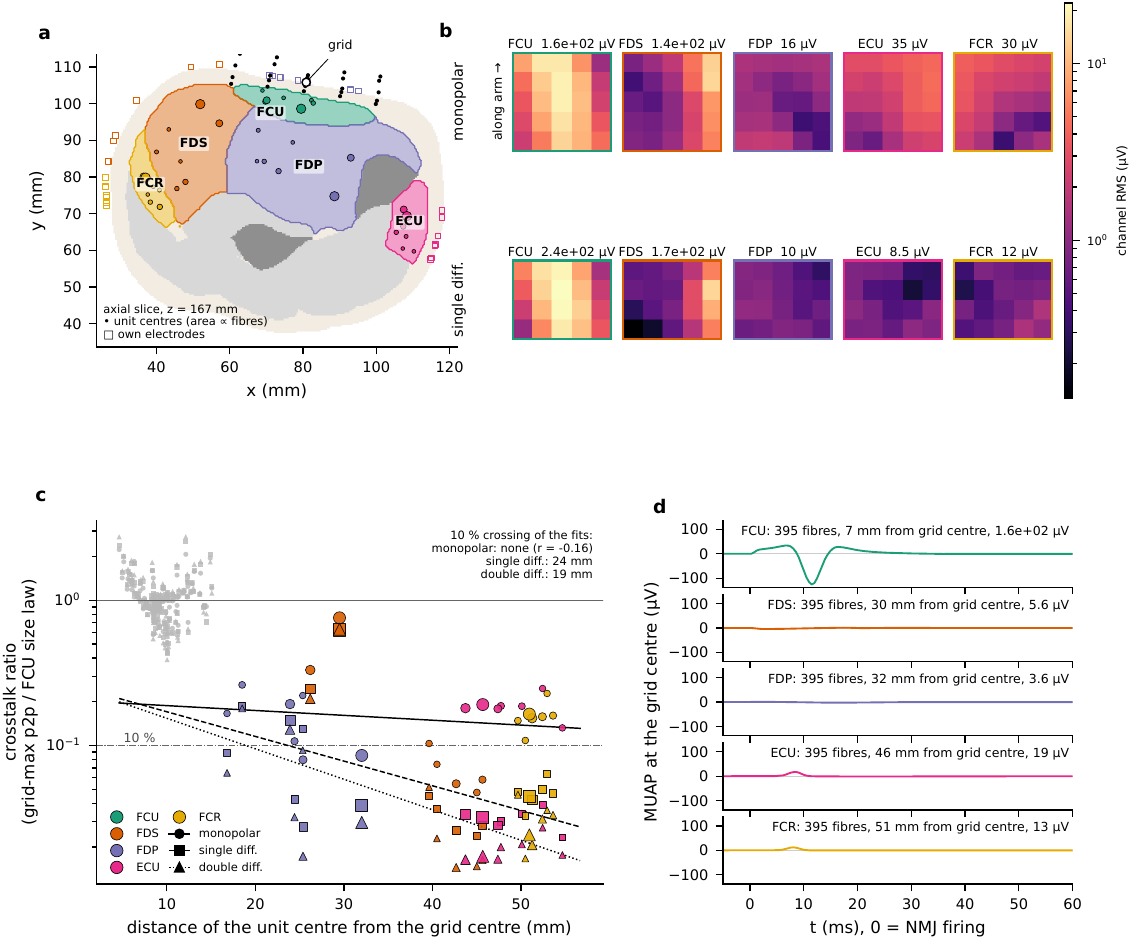}
\caption{Crosstalk on the FCU grid from the neighbouring muscles. (a) The forearm slice at
the grid centre: the FCU and its four neighbours, the seven unit centres per muscle (area
$\propto$ fibres), the grid, and one ``own'' skin electrode per unit (squares). (b)
Monopolar (top) and single-differential (bottom) RMS maps on the grid for the largest unit
($395$ fibres) of each muscle, shared log colour scale, peak-to-peak amplitude above each
map. (c) Ratio $R$ of each neighbouring unit's largest amplitude on the grid to the FCU
size law at its size, against distance from the grid centre, for the three montages
(log-linear fits and their $10\%$ crossings); grey: the FCU units themselves. (d) MUAPs at
the grid centre for the largest unit of each muscle at a shared scale, with each unit's
distance from the grid centre.}
\label{fig:crosstalk}
\end{figure}

The same $25$ lead fields serve every muscle in the segmentation, so this question
costs no new finite-element work. We built straight beds and size-principle pools for the
four muscles around the FCU --- flexor digitorum superficialis (FDS; lateral and
superficial, its edge $4$~mm from the grid's outer column), flexor digitorum profundus
(FDP; deep, directly beneath the FCU), extensor carpi ulnaris (ECU) and flexor carpi
radialis (FCR; on the far sides of the ulna and of the FDS) --- and synthesised seven units
per muscle ($5$--$395$ fibres) on the FCU grid (Fig.~\ref{fig:crosstalk}). Crosstalk is
expressed as the ratio $R$ of a neighbouring unit's largest amplitude on the grid to what
an FCU unit of the same size gives (the FCU size law of the 100-unit pool, $r = 0.97$).

Monopolar, $R$ does not fall with distance: it is $0.05$--$0.25$ for every unit from
$17$ to $55$~mm from the grid centre (fitted slope indistinguishable from zero,
$r = -0.16$); per muscle the medians are $0.07$ (FDS), $0.17$ (FDP), $0.19$ (ECU) and
$0.16$ (FCR), and $22$ of the $28$ units exceed $10\%$. What reaches the grid from the far
muscles is a single positive lobe at $\approx8$~ms --- the non-propagating end-of-fibre
far field --- which is why it neither walks along the grid nor fades with distance. The
largest ECU and FCR units, $46$--$51$~mm around the arm, put $35$ and
$30$~$\mu$V on the grid and $19$ and $13$~$\mu$V at its centre --- more than the largest
FDP unit directly beneath the FCU ($16$~$\mu$V on the grid, $3.6$ at the centre), because
depth attenuates a field faster than distance around a limb does. The FDS is a different
case: its largest unit's territory lies under the grid's outer column and its innervation
zone under the fourth row, so it reaches $0.76$ of the FCU value monopolar and $0.63$ in
either differential montage --- not crosstalk a montage can remove, but a grid that covers
two muscles. Differential montages
change the picture. Relative to monopolar, single-differential keeps a median $0.34$ of
the neighbours' amplitude ratio ($0.19$ of their energy) and double-differential $0.24$
($0.06$), while for the FCU's own units the same montages keep $1.0$ and $1.6$ of the
energy, so the own-to-crosstalk energy ratio improves $5.5\times$ and $28\times$; $R$ now
falls with distance ($r \approx -0.6$), crosses $10\%$ at $24$~mm
(single-differential) and $19$~mm (double-differential), and no unit beyond $30$~mm
exceeds it (monopolar: $55$~mm).

\emph{What to take from it.} A monopolar grid over a forearm muscle is not selective for
that muscle: units of a superficial neighbour $50$~mm away appear at $10$--$20\%$ of the
amplitude of an own unit, and distance alone does not remove them. Differentiation does,
at no cost to the own units --- the design rule is a differential montage, and a grid
narrow enough not to overlap the neighbour's territory. The pipeline gives these numbers
for any muscle and any layout in the segmentation for the price of one lead-field solve
per electrode. Two caveats: the monopolar potentials are referenced to the volume mean of
the conductor rather than to a remote electrode, so the monopolar far-field ratios are
reference-dependent (the differential ones are not); and all five muscles are cut by the
imaged volume, so the straight beds span the field of view rather than anatomical
fascicle lengths.

\subsection{Interference EMG across drive}
\label{sec:study-drive}
Six trapezoid contractions of the forearm pool at drives $0.1$--$1.0$ (dataset D3) give
plateau RMS of $14$, $27$, $47$, $73$, $101$ and $133$~$\mu$V at the centre electrode for
forces of $7$, $16$, $30$, $46$, $69$ and $100\%$~MVC, with $53$, $68$, $81$, $90$, $98$
and $100$ of the $100$ units active. The EMG--force relation is between linear and
quadratic (RMS at half force $0.56$ of that at full force), where
\citet{lawrence1983myoelectric} place the forearm muscles. The amplitude distribution
moves from super-Gaussian at low drive to Gaussian at high, as the number of
superimposed trains grows \citep{clancy1999probability}, and the median frequency at
moderate drive is $77$~Hz. Amplitude cancellation --- the loss of RMS when MUAPs of
opposite sign overlap --- is $68\%$ already at $20\%$ drive, twice the $\approx33\%$ that
\citet{keenan2005influence} report at that level; a movement trial modulates the RMS
$8.7\times$ between flexed and extended.

\emph{What to take from it.} The pool reproduces the shape of the EMG--force curve and
the statistics of the interference signal without tuning, so the drive can be read from
the RMS with the literature's calibration. Cancellation is the one statistic the pool
overshoots, and it tells us something about the fibre model: $100$ units firing $18$~ms
MUAPs from shared, full-length fibres overlap more than a real pool's do, because their
surface fields are broader than real units' --- a property of fibre length and the single
junction, not of curvature (Sec.~\ref{sec:study-fibres}). Absolute
RMS values carry the $\sigma_{\mathrm{in}}\pi a^{2}$ caveat of Sec.~\ref{sec:direct}
and should be compared within the model, not across to recordings.

\section{Discussion and outlook}
\label{sec:discussion}

\emgforge{} turns a segmentation into surface EMG with no hand steps, and every stage
in between is a checkable object. The two decisions that shape it are reciprocity, which
makes the cost proportional to electrodes rather than fibres, muscles or contractions,
and direct line-source synthesis, which evaluates the physics of the SFAP directly on the
lead field and is checked against a closed-form solution. The validation suite found the
model's own mistakes along the way --- a spurious $1/v$ in the amplitude, an electrode
placement that had been misread as a model property, and the anti-phase behaviour of our
spatial-frequency port --- and we list what remains open rather than smooth it over:

\begin{itemize}
\item \emph{Conduction velocity and amplitude.} The action potential is fixed in space,
  so amplitude is independent of $v$; physiologically it is fixed in time and a faster
  fibre carries a longer wave, so single-fibre amplitude should fall with $v$
  \citep{nandedkar1983simulation}.
\item \emph{Finite-element amplitude and depth law.} On FEM $\varphi$ the SFAP amplitude
  is erratic at $\pm40\%$ across depth because the recipe integrates $\varphi''$
  faithfully down to $\approx8$~mm wavelengths, where FEM $\varphi$ carries mesh structure
  the three-pole fit does not remove; low-passing $\varphi$ smooths the amplitudes but
  halves the mean frequency ($46$ vs $112$~Hz). The FEM cylinder also decays $\approx30\%$
  slower with depth than the analytical one. A finer mesh around the fibre band, and a
  refinement study of the $\approx47{,}000$-cell forearm mesh, are the next numerical work.
\item \emph{Fibre model.} Full-length, single-junction fibres --- straight or curved
  (Sec.~\ref{sec:study-fibres}) --- give surface fields broader than real units'; the largest units stay below the $1$--$2$~mV of the
  largest superficial recorded units and cancellation is overshot. The series-fibering
  variant of the harmonic bed addresses this and awaits validation.
\item \emph{Spatial-frequency port.} Anti-phase and lagged against the integral; to be
  audited for the action-potential orientation and sign conventions.
\item \emph{Pool.} The Gaussian renewal process has no refractory floor.
\item \emph{No experimental confrontation yet.} The next validation is spike-triggered
  surface MUAPs from decomposed HD-EMG on the same subject against the MRI pipeline, with
  the error and correlation targets of \citet{lowery2004volume, botelho2019anatomically}.
\end{itemize}

Three uses are immediate. The interference-EMG trials carry ground-truth spike trains,
unit properties and a full provenance record, which is what a decomposition benchmark
such as MUniverse \citep{mamidanna2025muniverse} consumes; an exporter to its BIDS-based
layout is the natural next step. The lead fields of the parametric limb are the training
data for neural-operator surrogates \citep{halatsis2026neural}, and the validation tiers
give those surrogates a signal-level target (the SFAP, not only the field). And a forward
model that is exact against first principles is the reference that model-informed
decomposition \citep{halatsis2025bmiss} needs when it inverts a learned MUAP generator.
Because the pipeline takes any segmentation, the same studies can be repeated on other
muscles, limbs and subjects without new code; that, more than any single number here, is
what it is for.

\section{Released datasets}
\label{sec:datasets}
Three datasets are generated by scripts in \texttt{paper/figures/} and described by
manifests in \texttt{paper/datasets/} (Table~\ref{tab:data}). They are reference
material: every array can be regenerated from the repository.

\begin{table}[t]
\centering\small
\caption{Datasets released with the paper (float32, NumPy \texttt{.npz} with a JSON
manifest each; $8$~MB in total).}
\label{tab:data}
\begin{tabular}{lp{86mm}r}
\toprule
name & contents & size \\
\midrule
\texttt{cylinder\_sfap\_atlas} & analytical and FEM lead fields and SFAPs on the validation cylinder for $6$ depths $\times$ $5$ angles $\times$ $3$ junction offsets $\times$ $9$ electrode positions ($1620$ SFAPs), with the Farina-generator MUAP of each case & 1.6~MB \\
\texttt{forearm\_fcu\_mu\_pool} & the 100-unit FCU pool (fibre paths, fibre indices, sizes, territories, activation parameters), MUAPs at the centre electrode and the full $(100, 25, 256)$ HD-grid MUAP tensor, electrode positions, configuration & 2.4~MB \\
\texttt{interference\_emg\_trials} & trapezoid contractions at six drive levels ($0.1$--$1.0$): drive, spike trains, single-channel and $25$-channel EMG, force; one movement trial & 4.1~MB \\
\bottomrule
\end{tabular}
\end{table}

\section*{Code and data availability}
Code: \repo{} (MIT), branch \path{paper/arxiv-v1}. The end-to-end runner is
\path{scripts/run_pipeline.py}; the validation suite is
\path{scripts/validation/run_all.py}; the figure, study and dataset scripts are in
\path{paper/figures/}; dataset manifests in \path{paper/datasets/}.

\section*{Funding}
D.H. was supported by UK Research and Innovation (UKRI Centre for Doctoral Training in AI
for Healthcare, grant EP/S023283/1), the Imperial-META Wearable Neural Interfaces Research
Centre and the Onassis Foundation (Scholarship ID F~ZT~012-1/2023-2024). N.E.-N. was
supported by UK Research and Innovation (UKRI AI Centre for Doctoral Training in Digital
Healthcare, grant EP/Y030974/1).

\FloatBarrier
\bibliographystyle{unsrtnat}
\bibliography{refs}

\appendix
\section{The validation checks}
\label{app:checks}
Table~\ref{tab:checks} lists every check of the suite with its criterion and the measured
result, generated from the suite's JSON records (\texttt{paper/make\_checks\_table.py}).
\small
\IfFileExists{checks_table.tex}{
\begin{longtable}{p{41mm}p{44mm}p{47mm}p{9mm}}
\caption{Every check of the validation suite: what it asserts, its criterion (first clause), what was measured (first clause), and the outcome. Full strings are in \texttt{docs/validation/REPORT.md}.}\label{tab:checks}\\
\toprule check & criterion & measured & result \\
\endfirsthead
\toprule check & criterion & measured & result \\
\endhead
\midrule
\multicolumn{4}{l}{\textit{Tier A --- cylinder: first principles, analytical, FEM, pipeline}} \\
\midrule
A0.1 spatial engine $\equiv$ first-principles line-source integral (shape, timing) & r $\geq$ 0.999 and |lag| $\leq$ 0.1 ms: same integral, so only discretisation can differ & r = [1.0, 0.9999, 0.9999], lag(ms) = [-0.0, -0.05, -0.05] over NMJ offsets 0/$-$20/$-$30 mm & pass \\
A0.2 spatial engine amplitude constant equals first principles & ratio 1.00 $\pm$ 5 \%, independent of v & engine/reference amplitude ratio = 0.994 at v=2, 0.994 at v=4 & pass \\
A0.3 Fourier engine vs first principles (signed r, lag) & r $\geq$ +0.99 (same integral, same sign, no lag) & signed r = [-0.795, -0.684, -0.643] at lag [1.4, 2.55, 0.85] ms;  vs a spatially MIRRORED IAP: r = [-0.788, -0.684, -0.642] & known \\
A0.4a sampling invariance: fsamp 4096$\to$2048 and dz 0.98$\to$0.5 mm leave the SFAP unchanged & r > 0.995, amplitude within 2 \% (discretisation-converged) & fsamp: r=1.00000 p2p ratio=0.9977;  dz: r=0.99886 p2p ratio=1.0011 & pass \\
A0.4b translation: moving the electrode $\Delta$z along the fibre delays the propagating lobe by $\Delta$z/v & |$\Delta$| $\leq$ one sample (0.24 ms) & $\Delta$z=20 mm $\to$ main-lobe delay 4.88 ms (expect 5.00) & pass \\
A0.5 the CSD source is monopole-free: $\int$ i\_m(z,t) dz = 0 at every instant & < 1e-6: generation and end-of-fibre terms exactly balance the propagating tripoles & max\_t |$\Sigma$\_z i\_m| / $\Sigma$\_z |i\_m| = 7.7e-17 (boxcar), 7.7e-17 (one\_sided) & pass \\
A0.4c superposition: a two-fibre MUAP is the sum of its SFAPs & < 1e-9 (linear volume conductor) & max|MUAP $-$ (SFAP$_1$+SFAP$_2$)| / p2p = 0.0e+00 & pass \\
A0.4d CV scaling: v 4$\to$3 stretches the SFAP by 4/3 and scales its spectrum by 3/4 & both within 8 \% & duration ratio 1.328 (expect 1.333), MNF ratio 0.757 (expect 0.750) & pass \\
A1.1 FEM $\varphi$(z) reproduces the analytical cylinder $\varphi$(z) along the fibre (5 depths) & r $\geq$ 0.99; FWHM within 20 \% (electrode models differ: $\varnothing$10 mm disk vs $\sigma$=5 mm Gaussian blob, so the shallowest fibres see\ldots{} & r = [0.9962, 0.9948, 0.9969, 0.9984, 0.9998]; FWHM\_z ana/FEM (mm) = ['41.0/35.1', '53.1/45.3', '63.1/57.7', '68.9/65.8', '81.2/82.1'] at r = 33/30/27/25/20 mm & pass \\
A1.1b \ldots{} and its second derivative $\varphi$''(z) --- the kernel the SFAP actually integrates & r $\geq$ 0.98 at the best-matched electrode radius (the FEM electrode is a $\sigma$=5 mm Gaussian blob clipped by the skin --- the analytical disk\ldots{} & r($\varphi$'') vs analytical electrode radius 2.5/5/7.5/10 mm: r=33: [0.985, 0.964, 0.916, 0.842]; r=27: [0.931, 0.923, 0.909, 0.889]; r=20: [1.0, 1.0, 1.0,\ldots{} & \textbf{fail} \\
A1.2 FEM/analytical SFAP amplitude ratio is one constant across depth (same depth law) & TWO findings. (i) A smooth 1.3--1.4$\times$ drift survives every preprocessing and a $\sigma$=1 mm source: the FEM cylinder decays $\sim$30 \% slower\ldots{} & p2p ratio FEM/ana, r=33$\to$20 mm, rel. to r=33 --- direct method: [1.0, 1.46, 1.35, 1.2, 0.86] (spread 1.69$\times$); no denoise: [1.0, 1.59, 1.3, 1.72, 2.03]\ldots{} & pass \\
A1.3 FEM $\varphi$(z) vs analytical for fibres off the electrode meridian ($\theta$ = 10--45$^\circ$) & r $\geq$ 0.99 & r = [0.9978, 0.9997, 0.9985, 0.997] & pass \\
A2.1 pipeline on FEM $\varphi$ $\equiv$ pipeline on analytical $\varphi$ (4 depths $\times$ 2 NMJ offsets) & r $\geq$ 0.95 with the direct recipe; deep fibres $\geq$ 0.99. The shallow-fibre residual is the FEM field (electrode model + mesh ripple through\ldots{} & direct method: r = [0.948, 0.97, 0.9, 0.929, 0.946, 0.915, 0.998, 0.995] (depth 33,33,30,30,25,25,20,20 mm; no denoise: [0.948, 0.954, 0.891, 0.909, 0.919,\ldots{} & known \\
A2.2 monopole denoise removes FEM mesh ripple without changing the analytical answer & denoised jaggedness $\leq$ raw; agreement not reduced & SFAP jaggedness raw$\to$denoised = ['0.004$\to$0.003', '0.007$\to$0.002', '0.006$\to$0.003', '0.011$\to$0.002'] \ldots{} & pass \\
A2.3 end-of-fibre onset lands at L/v in the FEM pipeline (and where the Farina generator puts it) & pipeline |$\Delta$| $\leq$ 0.5 ms & expected/Farina/pipeline (ms): 10.0/7.57/10.26, 15.0/12.45/15.39  $\to$ pipeline |$\Delta$| $\leq$ 0.39 ms; Farina onset leads by up to 2.55 ms (its IAP body\ldots{} & pass \\
A2.4 conduction velocity recovered from a 4-channel longitudinal array (both models) & within 5 \% of the set CV, linear lag--distance R$^2$ > 0.99 & Farina 3.83 m/s (R$^2$=0.998); FEM pipeline 3.98 m/s (R$^2$=0.999); set 4.0 & pass \\
A2.5 waveform |r| vs the Farina generator (sign-agnostic, $\pm$6 ms lag) & |r| $\geq$ 0.9 --- currently limited by the IAP-orientation/sign conventions of the Fourier/Farina family (A0.3) & spatial/Fourier: -0.91/+0.94, -0.74/+0.74, -0.93/+0.52 at NMJ 0/$-$20/$-$30 mm & known \\
A3.1 rotational symmetry: electrode +30$^\circ$ $\equiv$ fibre $-$30$^\circ$ & r > 0.999, peak within 3 \% (mesh-discretisation level) & r = [1.0, 1.0], peak ratio = [0.9986, 1.0109] & pass \\
A3.2 axial translation: electrode +20 mm $\equiv$ fibre window $-$20 mm (Neumann end effect) & r > 0.995, peak within 3 \% --- the residual ($\leq$0.5 \%) is the finite-length (240 mm) mesh's end effect & r = [0.99865, 0.99561], peak ratio = [0.9984, 0.9946] (electrode 100 mm from the mesh end) & pass \\
A3.3 reciprocity between two interior points in the anisotropic muscle & ratio 1 $\pm$ 5 \% (symmetric $\sigma$ $\Rightarrow$ Green's function symmetric) & $\varphi$\_A(B) = 2.3625e-02, $\varphi$\_B(A) = 2.3311e-02, ratio 1.0135 & pass \\
\midrule
\multicolumn{4}{l}{\textit{Tier B --- MUAP features against the literature}} \\
\midrule
B1 monopolar SFAP over the fibre (between IZ and tendon) is triphasic + $-$ + with the main lobe negative & + $-$ + with the largest lobe negative (depolarised zone under the electrode). The Farina port's propagating main lobe is POSITIVE: its output\ldots{} & pipeline lobes [1, -1, 1] (largest -1); Farina generator lobes [-1, 1, -1, 1, -1] (largest +1); electrode 30 mm from the NMJ, tendons $\geq$ 95 mm away,\ldots{} & pass \\
B2 conduction velocity recovered from an SD array tracks the set CV (3, 4, 5 m/s) & within 5 \% (pipeline) / 6 \% (Farina) at each CV; physiological range 3--5 m/s & pipeline/Farina: v=3: 3.13/3.15 m/s, v=4: 3.98/4.20 m/s, v=5: 4.88/nan m/s (Farina NaN = its k-grid does not build at that v) & pass \\
B3 innervation-zone signature: monopolar mirror symmetry, SD null and phase reversal at the IZ & mono r > 0.99; SD at the IZ < 5 \% of max; SD r(+d,$-$d) < $-$0.95 (mirror images of opposite sign) & pipeline: mono r(+d,$-$d) = [0.9981, 0.9979, 0.9985], SD@IZ/max = 0.041, SD r(+d,$-$d) = [-0.999, -0.998, -0.998];  Farina: SD@IZ/max = 0.004, SD r(+d,$-$d) =\ldots{} & pass \\
B4a the end-of-fibre component is non-propagating: same onset on every electrode & spread < 0.5 ms across electrodes $-$30\ldots{}+30 mm & pipeline onsets [10.51, 10.51, 10.51, 10.51, 10.51] ms (spread 0.00); Farina [7.57, 7.81, 7.57, 7.57, 7.57] (spread 0.24); expected L/v = 10 ms & pass \\
B4b EOF / propagating amplitude ratio increases monotonically with depth & monotone increase (the propagating part decays faster than the far-field EOF) & ratio at depth-below-skin 7/10/13/15/20 mm = [0.021, 0.053, 0.097, 0.127, 0.204] & pass \\
B4c spatial filters suppress the EOF: EOF/propagating ratio mono > SD > DD & strict ordering & \{'mono': 0.053, 'SD': 0.014, 'DD': 0.004\} (IED 10 mm) & pass \\
B5 amplitude vs depth is a power law (log-log linear), steeper for SD than monopolar & R$^2$ > 0.95 over 7--25 mm; n\_SD > n\_mono; monotone decrease. Note the Farina generator's own exponent differs from the first-principles engine's\ldots{} & exponent n (p2p $\propto$ d\^{}$-$n), R$^2$: pipeline mono 2.94 (0.998), SD 3.48 (0.998); Farina generator mono 2.04 (0.986); FEM pipeline mono 3.17 (0.987)\ldots{} & pass \\
B6 transverse spread: monopolar wider than SD (lit. 2.5--4$\times$); depth $\approx$ 0.2 $\times$ (mono 50\ldots{} & mono $\geq$ SD width; both widen with depth. The literature ratio (2.5--4$\times$; biceps MUs 15--25 mm deep, mono 72--96 mm vs SD 24--32 mm) comes\ldots{} & 50 \%-width (mm) at depth 15/20 mm: mono 45/88, SD 41/53 (ratio 1.1/1.7); depth/(0.2$\cdot$width) = 1.66/1.13 (Roeleveld's biceps-MU rule = 1.0; single fibre\ldots{} & pass \\
B7 MUAP amplitude $\propto$ fibre count at fixed geometry; NMJ scatter ($\sigma$=5 mm) disperses and\ldots{} & exact linearity; sub-linear amplitude and longer duration with dispersion & 50 identical fibres = 50$\times$SFAP to 9.9e-16; with NMJ $\sigma$=5 mm: p2p 0.86$\times$ the coherent sum, 10 \%-duration 20.0 $\to$ 22.2 ms & pass \\
B8 the bipolar montage is a pure spatial difference of a translation-invariant wave ($\Rightarrow$ comb\ldots{} & equivariance < 5e-3: SD(t) = S(t) $-$ S(t $-$ IED/v) to that precision, so |H| = |2 sin($\pi$ f$\cdot$IED/v)| with nulls at n$\cdot$v/IED follows\ldots{} & translation equivariance residual (single travelling wave, 9--33 ms): no denoise 1.4e-03, direct method 2.2e-03.  Measured null of |SD|/|mono| --- pipeline (no\ldots{} & pass \\
B9 electrode size is a spatial low-pass: amplitude and MNF fall with electrode area; $\varnothing$5 mm\ldots{} & monotone decrease; $-$3 dB points as in the literature & Farina p2p vs radius (0.5, 2.5, 5.0, 10.0): [1.0, 0.98, 0.924, 0.765]; MNF: [81.7, 81.0, 78.8, 72.8] Hz; disc filter: $\varnothing$5 mm @100 c/m = -2.8 dB,\ldots{} & pass \\
B10 SD amplitude grows $\approx$linearly with IED when IED $\ll$ $\lambda$, then saturates towards $\lambda$/2 & differentiator regime then saturation ($\lambda$ $\approx$ v$\cdot$T\_lobe $\approx$ 20--30 mm) & SD p2p vs IED (2.5, 5.0, 10.0, 20.0) mm --- pipeline: [1.0, 1.95, 3.59, 5.28]; Farina: [1.0, 1.95, 3.52, 5.01] ($\times$2 IED $\to$ $\geq$1.7$\times$ at small\ldots{} & pass \\
B11 subcutaneous fat attenuates (Kuiken 2003: $-$31/$-$80/$-$90 \% at 3/9/18 mm) and low-passes the surface\ldots{} & within $\times$1.5 of Kuiken's FE curve; RMS monotone decreasing and MNF lower at 3 and 9 mm than at 0.5 mm; FEM (smoothed $\varphi$) vs analytical\ldots{} & analytical RMS retained at fat (0.5, 3.0, 6.0, 9.0, 18.0) mm: [1.0, 0.498, 0.312, 0.226, 0.124] (Kuiken 0.69/0.20/0.10 at 3/9/18); MNF [91.0, 79.0, 76.0, 79.0,\ldots{} & pass \\
B12 muscle anisotropy ($\sigma$\_z/$\sigma$\_r = 5) elongates the lead field along the fibres & $\surd$5 exactly in the infinite medium; 1.3--3 in the layered cylinder (fat/skin layers and the bounded cross-section compress the elongation); FEM\ldots{} & FWHM\_z(aniso)/FWHM\_z(iso): analytical cylinder 1.48, infinite medium 2.236 (theory $\surd$5 = 2.236); FEM along/across FWHM at r=30: 2.29 (layered, bounded\ldots{} & pass \\
\midrule
\multicolumn{4}{l}{\textit{Tier C --- interference EMG and motor-unit pool}} \\
\midrule
C1 discharge rates lie in 5--40 pps and follow the onion skin (earlier-recruited fire faster) & model rates within 5--40 pps; measured within 10 \% of nominal; $\rho$(threshold, rate) < $-$0.9 & model: min rate 5, peak 40 pps; measured @drive 0.5: 89 active, 7.6--36.5 pps, Spearman(threshold, rate) = -0.98; @drive 1.0 max 42.9 pps (renewal-process\ldots{} & pass \\
C2 inter-spike-interval variability: CoV 0.1--0.3, no ISI below the $\sim$20 ms refractory floor & CoV $\in$ [0.1, 0.3]; < 1 \% of ISIs under 20 ms. The Gaussian renewal model has no refractory floor --- at 40 pps (ISI 25 ms, $\sigma$ 4 ms) sub-20\ldots{} & median per-MU ISI CoV 0.16 (model ISI\_CV = 0.167); ISIs < 20 ms: 1.58 \%; min ISI 9.8 ms & \textbf{fail} \\
C3 recruitment thresholds are right-skewed (many low, few high); twitch forces span $\approx$100$\times$ & skewness > 0.5; 50--200$\times$ (Fuglevand RP = 100; FDI 130) & threshold skewness +1.25 (median/max = 0.12); twitch P\_max/P\_min = 95 & pass \\
C4 amplitude PDF: super-Gaussian at low force, $\to$ Gaussian (kurtosis 3--3.5) at high; ARV/RMS $\in$ [0.71,\ldots{} & kurtosis decreasing with force to 2.8--3.6; ARV/RMS between Laplacian and Gaussian. NB: with the C-04 MUAPs (33 ms long) the low-force EMG is denser\ldots{} & kurtosis @drive 0.05/0.1/0.5/0.9 = [5.59, 4.56, 3.1, 2.92]; ARV/RMS @0.3/0.7 = [0.768, 0.786] (Laplacian 0.707, Gaussian 0.798) & pass \\
C5 amplitude cancellation grows with excitation ($\approx$30 \% at 20 \% $\to$ $\approx$60 \% at maximum) & increasing; 15--60 \% at 0.2, 40--85 \% at 1.0 (Keenan 2005: 33 \% $\to$ 62--65 \%). Cancellation is set by MUAP duration $\times$ active-MU count:\ldots{} & 1 $-$ mean|$\Sigma$ trains| / $\Sigma$ mean|train\_i| @drive 0.2/0.5/1.0 = [68, 79, 81] \% & known \\
C6 EMG--force relation lies between linear and quadratic: RMS at 50 \% force = 0.3--0.6 of RMS at 100 \% & 0.3 (quadratic, biceps/triceps) -- 0.6 (linear, FDI/soleus); force monotone in drive. A ratio above 0.6 (EMG concave in force) follows from the\ldots{} & RMS(50 \% MVC)/RMS(100 \% MVC) = 0.56; force plateau \%MVC = [7, 16, 30, 46, 69, 85, 100] at drive [0.1, 0.2, 0.35, 0.5, 0.7, 0.85, 1.0] & pass \\
C7 interference-EMG spectrum: MNF/MDF at moderate force in 70--130 Hz & MDF 70--130 Hz (bipolar norms: biceps 90$\pm$18, TA 116$\pm$20). Expected to fail while the C-04 fibre geometry yields 33 ms MUAPs --- the spectrum\ldots{} & @drive 0.5 (monopolar): MNF 81 Hz, MDF 77 Hz & pass \\
C8 larger motor units produce larger MUAPs (size $\leftrightarrow$ amplitude on the recording side) & $\rho$ > 0.5 (amplitude $\propto$ fibre count at fixed depth; depth scatter lowers $\rho$) & Spearman(fibres per MU, grid-max p2p) = +0.98 over 100 MUs (sizes 5--395 fibres) & pass \\
\midrule
\multicolumn{4}{l}{\textit{Tier S --- chain-level sanity}} \\
\midrule
Orderly recruitment (Henneman size principle) & n\_active non-decreasing in drive; smallest units recruited first & n\_active @ drive [0.05, 0.1, 0.2, 0.4, 0.6, 0.8, 1.0] = [37, 51, 67, 84, 94, 100, 100]; thresholds sorted=True & pass \\
Onion-skin rate coding & earlier (smaller) units discharge faster than later (larger) ones & @drive 0.7: first-recruited 40.0 Hz vs last 5.6 Hz (98 active) & pass \\
Interference-EMG amplitude $\uparrow$ with contraction & monotone increase with drive & plateau RMS($\mu$V) @ [0.1, 0.2, 0.35, 0.5, 0.7, 0.9] = [14.822, 28.279, 49.059, 61.24, 97.757, 124.795] & pass \\
EMG $\leftrightarrow$ force: monotone and MVC-calibrated & monotone in drive; $\approx$100 \%MVC at full drive & plateau force(\%MVC) @ [0.1, 0.2, 0.35, 0.5, 0.7, 0.9, 1.0] = [7.4, 16.0, 30.3, 46.8, 69.8, 92.2, 102.5] & pass \\
Conduction velocity from HD-EMG propagation & physiological muscle-fibre CV, mean $\approx$4 m/s --- the model uses one fibre CV ($\approx$uniform across the pool) & pool median CV 4.32 m/s (range 4.1--4.8 over 100 MUs, IED 10.1 mm); corr(size, CV) = -0.41 & pass \\
Spatial selectivity of a MUAP on the array & a MUAP is localised on the array, not a uniform far field (ratio > 1.5) & peak-channel RMS / grid-mean RMS = 2.16 (MU 99) & pass \\
Non-stationarity over a movement & EMG amplitude is modulated by joint angle, not just by the drive & RMS flexed / RMS extended = 8.66  (90.856 vs 10.495 $\mu$V) & pass \\
MUAP physiological scale & single-MU surface MUAP $\approx$ 20--800 $\mu$V, 5--20 ms (Merletti \& Muceli 2019; Farina 2014) (the 0.6 $\mu$V / 33 ms of earlier runs was an\ldots{} & median p2p 23.60 $\mu$V, median duration 18.1 ms & pass \\
\bottomrule
\end{longtable}
}{\emph{(table not generated yet)}}
\normalsize

\end{document}